\documentclass[aps,prd,singlecolumn,showpacs,superscriptaddress,groupedaddress,nofootinbib]{revtex4}
\usepackage{xcolor}
\usepackage{graphicx,color,bm}

\usepackage{amsmath,amssymb}
\usepackage{epstopdf}
\usepackage{ulem}
\usepackage{array}
\usepackage{verbatim}
\usepackage[utf8]{inputenc}
\usepackage{rotating}
\newcolumntype{K}[1]{>{\centering\arraybackslash}m{#1}}

\usepackage[colorlinks,linkcolor=blue,citecolor=blue]{hyperref}

\newcommand{\orcid}[1]{\href{https://orcid.org/#1}{\,\includegraphics[width=8px]{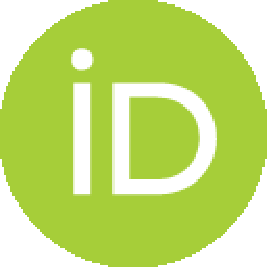}}}

\usepackage{mathtools}

\usepackage{tabularx}

\begin{document}

\title{A comprehensive data-driven odyssey to explore the equation of state of dark energy}


\author{Bikash R. Dinda\orcid{0000-0001-5432-667X}}
\email{bikashrdinda@gmail.com}
\affiliation{Department of Physical Sciences, Indian Institute of Science Education and Research Kolkata, Mohanpur, Nadia, West Bengal 741246, India.}
\affiliation{Department of Physics $\&$ Astronomy, University of the Western Cape, Cape Town, 7535, South Africa}

\author{Narayan Banerjee \orcid{0000-0002-9799-2813}}
\email{narayan@iiserkol.ac.in}
\affiliation{Department of Physical Sciences, Indian Institute of Science Education and Research Kolkata, Mohanpur, Nadia, West Bengal 741246, India.}

\begin{abstract}
For the first time, we reconstruct the dark energy equation of the state parameter $w$ from the combination of background and perturbation observations, specifically combining the Hubble parameter data from cosmic chronometer observations and the logarithmic growth rate data from the growth rate observations. We do this analysis using posterior Gaussian process regression without considering any specific cosmological model or parametrization. However there are three main assumptions: (I) a flat Friedmann-Lemaître-Robertson-Walker (FLRW) metric is considered for the cosmological background, (II) there is no interaction between dark energy and matter sectors, and (III) for the growth of inhomogeneity, sub-Hubble approximation and linear perturbations are considered. This study is unique in the sense that the reconstruction of $w$ is independent of any derived parameters such as the present values of the matter-energy density parameter and Hubble parameter. From the reconstruction, we look at how the dark energy equation of state evolves between redshifts 0 and 1.5, finding a slight hint of dynamical behavior in dark energy. However, the evidence is not significant. We also find a leaning towards non-phantom behavior over phantom behavior. We observe that the $\Lambda$CDM model ($w=-1$) nearly touches the lower boundary of the 1$\sigma$ confidence region in the redshift range $0.6 \lesssim z \lesssim 0.85$. However, it comfortably resides within the 2$\sigma$ confidence region in the whole redshift range under investigation, $0\leq z \leq 1.5$. Consequently, the non-parametric, model-independent reconstruction of dark energy provides no compelling evidence to deviate from the $\Lambda$CDM model when considering cosmic chronometer and growth rate observations.
\end{abstract}


\maketitle
\date{\today}

\section{Introduction}

Since 1998 from the observations of type Ia supernovae \citep{SupernovaCosmologyProject:1997zqe,SupernovaSearchTeam:1998fmf,SupernovaCosmologyProject:1998vns,2011NatPh...7Q.833W}, we have evidence for late-time cosmic acceleration and it has been confirmed by several other observations such as cosmic microwave background (CMB) \citep{Planck:2013pxb,Planck:2015fie,Planck:2018vyg}, baryon acoustic oscillation (BAO) \citep{BOSS:2016wmc,eBOSS:2020yzd,Hou:2020rse}, cosmic chronometers (CC) \citep{Jimenez:2001gg,Pinho:2018unz,Cao:2023eja} observations. This cosmic acceleration is explained by two broad kinds of models. One is the introduction of dark energy as a constituent of the Universe \citep{Peebles:2002gy,Copeland:2006wr,Yoo:2012ug,Lonappan:2017lzt,Dinda:2017swh,Dinda:2018uwm,Bamba:2012cp}. The second class of models is the modification of gravity \citep{Clifton:2011jh,Koyama:2015vza,Tsujikawa:2010zza,Joyce:2016vqv,Dinda:2017lpz,Dinda:2018eyt,Zhang:2020qkd,Dinda:2022ixi,Bassi:2023vaq,Silvestri:2009hh,Nojiri:2010wj,Nojiri:2006ri,Nojiri:2017ncd,deHaro:2023lbq}.

There have been numerous efforts to model dark energy \citep{Huterer:2017buf,Motta:2021hvl,Li:2012dt} and many of these models have been tested against observational data \citep{Lonappan:2017lzt}. The simplest and widely accepted model of dark energy is the $\Lambda$CDM model, in which the cosmological constant $\Lambda$ is considered to be the candidate for the dark energy \citep{Carroll:2000fy}. The $\Lambda$CDM model is the most successful in the sense that it has been tested up to a certain level against several cosmological observations, mentioned earlier. However, this model still has shortcomings. For example, this model encounters challenges such as the fine-tuning problem, where the theoretical vacuum energy density, derived from quantum field theory, significantly exceeds the observed values obtained from various cosmological observations \citep{Zlatev:1998tr,Sahni:1999gb,Velten:2014nra,Malquarti:2003hn}. This model is also associated with the cosmic coincidence problem, which questions why the current ratio of the cosmological constant to the total matter of the Universe is approximately 7:3 \citep{Zlatev:1998tr,Sahni:1999gb,Velten:2014nra,Malquarti:2003hn}. Also, in the $\Lambda$CDM model, there is a discrepancy in the observed values of $H_0$ (present value of the Hubble parameter) between early Universe observations \citep{DiValentino:2021izs,Krishnan:2021dyb,Vagnozzi:2019ezj,Dinda:2021ffa} like CMB \citep{Planck:2013pxb,Planck:2015fie,Planck:2018vyg} and late-time observations such as SHOES \citep{Riess:2016jrr,Riess:2020fzl} in local distance ladders. This is commonly referred to as the Hubble tension. Furthermore, in the $\Lambda$CDM model, there exists another observational inconsistency in the observed values of the matter density distribution fluctuations, quantified by the quantity $\sigma_8$, between early and late-time observations. This is commonly referred to as the $\sigma_8$ tension \citep{DiValentino:2020vvd,Abdalla:2022yfr,Douspis:2018xlj,Bhattacharyya:2018fwb}. This has motivated us to go beyond the $\Lambda$CDM model.

Various alternative dark energy models have been proposed, including quintessence \citep{Dinda:2016ibo} and k-essence \citep{Dinda:2023mad} models, as well as dark energy parametrizations such as wCDM \citep{Anselmi:2014nya}, Chevallier-Polarski-Linder (CPL) \citep{Chevallier:2000qy,Linder:2002et}, and Barboza-Alcaniz (BA) \citep{Barboza:2008rh} parametrizations. Each model has its limitations. For instance, the wCDM parametrization assumes a constant equation of state for dark energy, lacking a dynamical evolution \citep{Anselmi:2014nya}. Quintessence models exhibit evolving equations of state for dark energy but are confined to the non-phantom ($w>-1$) regions \citep{Dinda:2016ibo}. Notably, in the context of the Hubble tension, quintessence models worsen the tension, as it is worse for the non-phantom equation of state, while the phantom equation of state alleviates this tension comparatively up to a certain extent \citep{Banerjee:2020xcn}. Alongside these limitations, adopting any specific model or parametrization for dark energy may introduce potential biases in the results.

To avoid these limitations, the model-independent analysis is important to study the behavior of a dark energy component. In the literature, there are different approaches to study the behavior of dark energy through different model-independent analyses \citep{Mehrabi:2022ywh,AlbertoVazquez:2012ofj,Liu:2015mkm,Gerardi:2019obr,Bonilla:2020wbn,Sahni:2006pa,Mukherjee:2020ytg,Rezaei_2020,Capozziello:2022uak,Luongo:2015zgq,Wang_2009,Dinda:2019mev,Raveri:2021dbu,Pogosian:2021mcs,Mu:2023zct,Ruiz-Zapatero:2022zpx,Bernardo:2021cxi,Calderon:2022cfj}. For example, the dark energy has been investigated using cosmographic or kinematic analysis through cosmological quantities like the Hubble parameter, deceleration parameter, and the jerk parameter \citep{Rezaei_2020,Capozziello:2022uak,Luongo:2015zgq,Wang_2009}. Also, it has been studied through model-independent parametrizations of cosmological quantities like the Hubble parameter, comoving distance, luminosity distance through series expansions like Taylor series and  Pade$^\prime$ series expansion \citep{Dinda:2019mev}. So these kinds of studies are parametrized. Although these parametrizations are independent of any cosmological model, since these are parametrized, there might be still the presence of bias in the results. To get bias-free results, we should consider not only model-independent but also non-parametric approaches which are completely data-driven to study the behavior of dark energy \citep{Perenon:2022fgw}.

While numerous efforts have focused on reconstructing the equation of state of dark energy, many of these model-independent studies primarily rely on cosmological background observations \citep{Mehrabi:2022ywh,Liu:2015mkm,Gerardi:2019obr,Bonilla:2020wbn,Holsclaw_2011,Lazkoz_2012,Wang:2018fng,Teng:2021cvy}. However, a complete data-driven approach to understand the behavior of dark energy is incomplete when restricted to background cosmological observations alone. This limitation arises because background observations only offer information about the total energy budget of the Universe, not the individual components. When relying solely on background observations, reconstruction of the equation of the state of dark energy depends on additional derived cosmological parameters like the matter-energy density and the Hubble parameter at present \citep{Perenon:2022fgw}. Knowledge about these additional parameters may still require consideration of any cosmological model or parametrization.

On the other hand, there are very few prior studies which include perturbation observations to reconstruct the equation of state of dark energy. To the best of our knowledge, only one reference Perenon et al. \citep{Perenon:2022fgw} considers background (cosmic chronometer observations) and perturbation (growth rate observations) observations separately to reconstruct the equation of state of dark energy in a model-independent way. When relying solely on perturbation observations, similar to background observations, we still require prior information on the same cosmological parameters such as the matter-energy density parameter and the Hubble parameter at present \citep{Perenon:2022fgw}. However, by considering both background and perturbation observations together, we demonstrate, for the first time, that we can reconstruct the equation of state of dark energy entirely independent of any prior knowledge of these cosmological parameters. This novel approach arises from the fact that dark energy and matter play distinct roles in the process of structure formation \citep{Sahni:2006pa,Zhang:2018gjb,Wang:2007fsa,Ruiz:2014hma,Bernal:2015zom,DAgostino:2023cgx}. For this reason, we combine background and perturbation observations, specifically, cosmic chronometer data for the Hubble parameter \citep{Jimenez:2001gg,Pinho:2018unz,Cao:2023eja}, and growth rate data for the logarithmic growth rate \citep{Avila:2022xad} to explore the behavior of dark energy by studying the redshift evolution of its equation of state parameter.

In our study, for the model-independent methodology, we employ a powerful non-parametric approach known as the posterior Gaussian Process Regression (GPR) analysis \citep{williams1995gaussian,GpRasWil,Seikel_2012,Shafieloo_2012,Hwang:2022hla,Keeley:2020aym,Dinda:2022vmb,Dinda:2022jih,Perenon:2022fgw,Perenon:2021uom,OColgain:2021pyh,Banerjee:2023evd,Mukherjee:2023lqr,Banerjee:2023rvg,Mukherjee:2022ujw,Mukherjee:2020vkx,Mukherjee:2020ytg,Zheng:2023yco,Oliveira:2023uid}. This sophisticated technique serves as a valuable tool for the prediction of a smooth function from given data, utilizing a kernel covariance function and a mean function. Notably, GPR is useful to predict the first-order (and higher-orders too) derivative of the same function from the same data itself. We select a zero mean function, a widely favored choice in GPR analysis, to make our analysis entirely independent of any specific cosmological model \citep{williams1995gaussian,GpRasWil,Seikel_2012,Shafieloo_2012}. Using GPR, we reconstruct smooth functions of the Hubble parameter and its derivative from the cosmic chronometer data. Similarly, we reconstruct smooth functions of the logarithmic growth rate and its derivative from the growth rate data. From these, we reconstruct the equation of the state parameter of dark energy using an interesting equation, which is derived from the flat Friedmann-Lemaître-Robertson-Walker (FLRW) metric and Newtonian perturbation theory on linear scales. We compare this reconstructed result with the $\Lambda$CDM model.

Our analysis reveals that the reconstruction errors for the equation of state parameter, $w$, are notably narrower within the redshift range $0.5 \lesssim z \lesssim 1$, indicating tighter constraints on $w$ within this interval. Comparing the reconstructed $w$ values with those predicted by the standard $\Lambda$CDM model, we note that the $\Lambda$CDM model aligns closely with the lower boundary of the 1$\sigma$ reconstructed region within the redshift interval $0.6 \lesssim z \lesssim 0.85$ while remaining well within the 1$\sigma$ region across other redshift regions. Our analysis suggests a preference for dark energy exhibiting non-phantom behavior ($w > -1$). Although indications of its dynamical evolution are observed, this evidence is not significant. Our comprehensive non-parametric, model-independent study utilizing cosmic chronometer and growth rate observations yields no compelling evidence to deviate from the standard $\Lambda$CDM model.

In this paper, all the equations are written in the natural units. This paper is organized as the following: In Sec.~\ref{sec-eos_calculation}, we derive the expression for the equation of state of dark energy w.r.t the Hubble parameter, the growth rate, and their first derivatives w.r.t redshift, where we address the shortcomings of the exclusive use of the either of the background and perturbation observations. In Sec.~\ref{sec-errorpropagation}, we calculate the equation for the propagation of errors in the dark energy equation of state. In Sec.~\ref{sec-data}, we briefly discuss the cosmic chronometers and growth rate data. In Sec.~\ref{sec-method_and_result}, we discuss the methodology of GPR analysis to compute the Hubble parameter and the logarithmic growth rate and their derivatives and show the results by computing the redshift evolution of the dark energy equation of state. Finally, in Sec.~\ref{sec-conclusion}, we present a conclusion.

\section{Equation of state of dark energy}
\label{sec-eos_calculation}

\subsection{Exclusive use of background observations}

We consider flat Friedmann-Lema\^itre-Robertson-Walker (FLRW) metric for the background expansion of the Universe. Corresponding to this metric, the first Friedmann equation is given as

\begin{equation}
3 M_{\rm pl}^2 H^2 = \bar{\rho}_{\rm tot} = \bar{\rho}_m+\bar{\rho}_{\rm DE},
\label{eq:first_Friedmann}
\end{equation}

\noindent
where $M_{\rm pl}$ is the reduced Planck mass defined as $M_{\rm pl}^2=\frac{1}{8\pi G}$ with $G$ being the Newtonian gravitational constant, $H$ is the Hubble parameter, $\bar{\rho}_{\rm tot}$ is the total background energy density of the Universe, $\bar{\rho}_m$ and $\bar{\rho}_{\rm DE}$ are the background energy densities of the total matter (cold dark matter and baryon together) and dark energy respectively. In the above equation, in the second equality, we have considered that the Universe is dominated by matter and dark energy only and neglected the contribution from the radiation. In this case, the second Friedmann equation is given as

\begin{equation}
6 M_{\rm pl}^2 (\dot{H}+H^2) = -(\bar{\rho}_{\rm tot}+3 \bar{P}_{\rm tot}) = -(\bar{\rho}_m+\bar{\rho}_{\rm DE}+3 \bar{P}_{\rm DE}) = -\bar{\rho}_m-(1+3 w)\bar{\rho}_{\rm DE},
\label{eq:second_Friedmann}
\end{equation}

\noindent
where $\bar{P}_{\rm tot}$ is the total background pressure which is the background pressure contribution from the dark energy only (denoted by $\bar{P}_{\rm DE}$) because matter is assumed to be pressureless; $w$ is the equation of state of the dark energy defined as $\bar{P}_{\rm DE}=w \bar{\rho}_{\rm DE}$; overhead dot represents the differentiation w.r.t the cosmic time $t$. The matter-energy density parameter is defined as

\begin{equation}
\Omega_m = \frac{\bar{\rho}_m}{\bar{\rho}_m+\bar{\rho}_{\rm DE}}.
\label{eq:defnOmegaM}
\end{equation}

\noindent
Using Eqs.~\eqref{eq:first_Friedmann} and~\eqref{eq:defnOmegaM}, we can write $\bar{\rho}_m$ and $\bar{\rho}_{\rm DE}$ w.r.t $H$ and $\Omega_m$ given as

\begin{eqnarray}
\bar{\rho}_m &=& 3 M_{\rm pl}^2 H^2 \Omega_m,
\label{eq:sol_rho_m} \\
\bar{\rho}_{\rm DE} &=& 3 M_{\rm pl}^2 H^2 (1-\Omega_m).
\label{eq:sol_rho_de}
\end{eqnarray}

\noindent
Putting Eqs.~\eqref{eq:sol_rho_m} and~\eqref{eq:sol_rho_de} in Eq.~\eqref{eq:second_Friedmann}, we rewrite the second Friedmann equation given as

\begin{equation}
\frac{2(\dot{H}+H^2)}{H^2} = -1-3(1-\Omega_m)w.
\label{eq:second_FReqn_again_pre}
\end{equation}

\noindent
The differential equations w.r.t $t$ can be rewritten w.r.t the redshift $z$ using the relation

\begin{equation}
\dot{A}=-(1+z)HA',
\label{eq:dot_prime_conversion}
\end{equation}

\noindent
for any quantity $A$, where the overhead prime denotes the differentiation with the redshift $z$ as the argument. Using above relation, Eq.~\eqref{eq:second_FReqn_again_pre} can be rewritten as

\begin{equation}
3 + 3(1-\Omega_m)w - 2 (1+z) \frac{H'}{H} = 0 .
\label{eq:second_FReqn_again}
\end{equation}

\noindent
Eq.~\eqref{eq:second_FReqn_again} suggests that we can either determine $\Omega_m$ when $w$ is known or $w$ when $\Omega_m$ is known. In the first case, we have to consider any cosmological model or parametrization to the equation of state of dark energy so that we know $w$ as a function of redshift. In this case, using evolutions of $w$, $H$, and $H'$, we can compute the evolution of $\Omega_m$ by the equation given as

\begin{equation}
\Omega_m = 1 - \frac{ 2(1+z)H'-3H }{ 3wH } .
\label{eq:sol_Om_only_bkg}
\end{equation}

\noindent
Clearly, this is not the case in the current analysis, because our aim is not to use any cosmological model or parametrization.

On the other hand, if we somehow know the evolution of $\Omega_m$, we can compute the evolution of the equation of state of dark energy by the equation given as

\begin{equation}
w = \frac{ 2(1+z)H'-3H }{ 3H(1-\Omega_m) } .
\label{eq:sol_w_only_bkg}
\end{equation}

\noindent
From Eq.~\eqref{eq:sol_w_only_bkg}, we can see that to find $w$ exclusively from background analysis, we need prior knowledge of $\Omega_m$. The $\Omega_m$ is a derived variable i.e. we can not directly obtain its values from the observational data. Determination of its values depends on the additional cosmological parameters such as $\Omega_{\rm m0}$ (present value of $\Omega_m$) and $H_0$ (present value of $H$). For example, if we consider no interaction between dark energy and matter sectors, we can have prior knowledge of $\Omega_m$ from the prior knowledge of $\Omega_{\rm m0}$ and $H_0$ parameters using equation given as

\begin{equation}
\Omega_m = \frac{\Omega_{\rm m0}H_0^2(1+z)^3}{H^2} .
\label{eq:Om_wrt_Om0_H0}
\end{equation}

\noindent
For details, see Perenon et al. \citep{Perenon:2022fgw}. Furthermore, if we consider interaction between dark energy and matter sectors, it can depend on other parameters too. Note that, in this analysis, we restrict ourselves in the scenario where there is no interaction between dark energy and matter sectors. Thus non-parametric model independent determination of $w$ is not possible exclusively from the background data without any prior knowledge of parameters such as $\Omega_{\rm m0}$ and $H_0$. Further, determination of these parameters from the observational data may depend on any model or parametrization which we shall avoid.

\subsection{Exclusive use of perturbation observations}

The differential equation for growth of matter inhomogeneity $\delta_m$ is given as \citep{Dinda:2018uwm,Dinda:2023mad,Dinda:2017swh,Dinda:2018ojk,Dinda:2023kvg,Dinda:2023xqx}

\begin{equation}
\ddot{\delta}_m+2H\dot{\delta}_m-4\pi G \bar{\rho}_m \delta_m=0,
\label{eq:basic_matter_inhomogeneity}
\end{equation}

\noindent
where the overhead double dot represents the double differentiation w.r.t $t$. The above equation involves three major assumptions: (I) a flat FLRW metric is considered for the cosmological background, (II) there is no interaction between dark energy and matter sectors, and (III) for the growth of matter inhomogeneity, sub-Hubble approximation (where Newtonian perturbation theory is applicable) and linear perturbations (applicable at linear scales only) are considered. The logarithmic growth rate $f$ is defined as

\begin{equation}
f = \dfrac{d \ln \delta_m}{d \ln a} = \frac{\dot{\delta}_m}{H\delta_m},
\label{eq:defn_f}
\end{equation}

\noindent
where $a$ is the cosmic scale factor. This equation can be rewritten as

\begin{equation}
\dot{\delta}_m = H \delta_m f,
\label{eq:delta_m_dot}
\end{equation}

\noindent
which on differentiation, yields

\begin{equation}
\ddot{\delta}_m = \left( H\dot{f} +H^2f^2+\dot{H} f \right) \delta_m.
\label{eq:delta_m_double_dot}
\end{equation}

\noindent
Putting Eqs.~\eqref{eq:delta_m_dot},~\eqref{eq:delta_m_double_dot}, and~\eqref{eq:sol_rho_m} in Eq.~\eqref{eq:basic_matter_inhomogeneity}, we get a differential equation for $f$ given as

\begin{equation}
\dot{f}+\left( 2H+\frac{\dot{H}}{H} \right)f+Hf^2-\frac{3}{2}H \Omega_m=0.
\label{eq:f_eqn_wrt_t}
\end{equation}

\noindent
We rewrite the above differential equation, by using the relation in Eq.~\eqref{eq:dot_prime_conversion}, given as

\begin{equation}
-(1+z)f'+\left[ 2-(1+z)\frac{H'}{H} \right]f+f^2-\frac{3}{2} \Omega_m=0.
\label{eq:f_eqn_wrt_z}
\end{equation}

\noindent
From Eq.~\eqref{eq:second_FReqn_again}, we get

\begin{equation}
(1+z) \frac{H'}{H} = \frac{3}{2}\left[1 + (1-\Omega_m)w\right] .
\label{eq:second_FReqn_again_Hpfct}
\end{equation}

\noindent
Putting Eq.~\eqref{eq:second_FReqn_again_Hpfct} in Eq.~\eqref{eq:f_eqn_wrt_z}, one can write

\begin{equation}
-(1+z)f'+ \frac{1}{2} \left[ 1-3(1-\Omega_m)w \right]f+f^2-\frac{3}{2} \Omega_m=0.
\label{eq:f_eqn_wrt_z_Again}
\end{equation}

\noindent
From Eq.~\eqref{eq:f_eqn_wrt_z_Again}, we can either determine $\Omega_m$ if $w$ is known or $w$ if $\Omega_m$ is known. Thus we have

\begin{eqnarray}
\Omega_m &=& \frac{2 (1+z) f'+(3 w-1)f-2f^2}{3 (wf-1)} ,
\label{eq:sol_Om_only_prtn} \\
&& \nonumber\\
&& \hspace{2 cm} \text{or} \nonumber\\
&& \nonumber\\
w &=& \frac{2 f^2+f-2 (1+z) f'-3 \Omega_m}{3 (1-\Omega_m)f } .
\label{eq:sol_w_only_prtn}
\end{eqnarray}

\noindent
Similar to the previous case in Eq.~\eqref{eq:sol_w_only_bkg}, here also, from Eq.~\eqref{eq:sol_w_only_prtn}, we need prior information on the derived variable $\Omega_m$ to determine $w$ \citep{Perenon:2022fgw}. This makes it obvious that, exclusively from perturbation data, we can not determine $w$ in a non-parametric model-independent way.

\subsection{From the combination of background and perturbation observations}

Combining Eqs.~\eqref{eq:second_FReqn_again} and~\eqref{eq:f_eqn_wrt_z_Again}, we can write expression of $\Omega_m$ independent of $w$ given as

\begin{equation}
\Omega_m = -\frac{2}{3}(1+z)f' + \frac{2}{3} \left[ \left( 2-\frac{1+z}{H} H' \right)f+f^2 \right] ,
\label{eq:sol_Om_final}
\end{equation}

\noindent
and simultaneously, as well as expression of $w$ independent of $\Omega_m$ given as

\begin{equation}
w = \frac{2 (1+z) H'-3 H}{H \left[ 2 (1+z) f'-2 f (f+2)+3 \right] + 2 (1+z) H' f} .
\label{eq:final_sol_w}
\end{equation}

\noindent
Unlike in the previous two cases in Eqs.~\eqref{eq:sol_w_only_bkg} and~\eqref{eq:sol_w_only_prtn}, here in Eq.~\eqref{eq:final_sol_w}, we can see that the determination of $w$ is independent of the determination of $\Omega_m$. Thus, the reconstruction of $w$ is independent of any additional derived parameters such as $\Omega_{\rm m0}$ and $H_0$. So, in principle, from the combination of background (through $H$ and its derivative) and the perturbation (through $f$ and its derivative) data, it is possible to reconstruct the equation of the state of dark energy completely in a non-parametric model-independent way.

\section{Propagation of errors}
\label{sec-errorpropagation}

We compute error in $w$ from the errors of $H$, $H'$, $f$, and $f'$ using propagation of uncertainty given as

\begin{eqnarray}
\text{Var}[w] &=& \left( \frac{\partial w}{\partial H} \right)^2 \text{Var}[H] + \left( \frac{\partial w}{\partial H'} \right)^2 \text{Var}[H'] + 2 \frac{\partial w}{\partial H} \frac{\partial w}{\partial H'} \text{Cov}[H,H']  \nonumber\\
&& + \left( \frac{\partial w}{\partial f} \right)^2 \text{Var}[f] + \left( \frac{\partial w}{\partial f'} \right)^2 \text{Var}[f'] + 2 \frac{\partial w}{\partial f} \frac{\partial w}{\partial f'} \text{Cov}[f,f'],
\label{eq:var_w}
\end{eqnarray}

\noindent
where we have

\begin{align}
\frac{\partial w}{\partial H} &= \frac{2 (z+1) H' \left(2 f^2-2 (z+1) f'+f-3\right)}{\left(H \left(-2 (z+1) f'+2 f (f+2)-3\right)-2 f (z+1) H'\right)^2}, \nonumber\\
\frac{\partial w}{\partial H'} &= \frac{2 H (z+1) \left(2 (z+1) f'-f (2 f+1)+3\right)}{\left(H \left(-2 (z+1) f'+2 f (f+2)-3\right)-2 f (z+1) H'\right)^2}, \nonumber\\
\frac{\partial w}{\partial f} &= -\frac{2 \left(3 H-2 (z+1) H'\right) \left(2 (f+1) H-(z+1) H'\right)}{\left(H \left(-2 (z+1) f'+2 f (f+2)-3\right)-2 f (z+1) H'\right)^2}, \nonumber\\
\frac{\partial w}{\partial f'} &= -\frac{2 H (z+1) \left(2 (z+1) H'-3 H\right)}{\left(H \left(2 (z+1) f'-2 f (f+2)+3\right)+2 f (z+1) H'\right)^2}. \nonumber\\
&
\end{align}

\noindent
In the above expressions, $\text{Var}[A]$ represents the variance in any quantity $A$, and $\text{Cov}[A,B]$ represents covariance between any two quantities $A$ and $B$. In Eq.~\eqref{eq:var_w}, we have assumed no correlation between $H$ (or $H'$) and $f$ (or $f'$), because $H$ (or $H'$) and $f$ (or $f'$) are obtained from two different kinds of observational data, discussed in the next section.

\section{Observational data}
\label{sec-data}

We consider the cosmic chronometer observations consisting of 32 Hubble parameter data at various redshift values in the range $0.07 \leq z \leq 1.965$. For this data, we follow Cao and Ratra \cite{Cao:2023eja}. Among these 32 Hubble parameter data, 15 data are correlated to each other and we include these non-zero covariances too in our analysis\footnote{The covariances can be found in \url{https://gitlab.com/mmoresco/CCcovariance/} \citep{Moresco:2020fbm,moresco2012improved,Moresco:2015cya,Moresco:2016mzx}.}. The mean values and the standard deviations of the observed Hubble parameter are listed in Table~\ref{table:H_CC_data}. The last 15 redshift points with asterisks have non-zero covariances.

\begin{table}
\begin{center}
\begin{tabular}{|c|c|c|c|}
\hline
Sr. &$z$ & $H$ [km s$^{-1}$ Mpc$^{-1}$] & Refs. \\
\hline
1 & 0.07 & 69.0 $\pm$ 19.6 & \citep{zhang2014four} \\
2 & 0.09 & 69.0 $\pm$ 12.0 & \citep{Simon:2004tf} \\
3 & 0.12 & 68.6 $\pm$ 26.2 & \citep{zhang2014four} \\
4 & 0.17 & 83.0 $\pm$ 8.0 & \citep{Simon:2004tf} \\
5 & 0.2 & 72.9 $\pm$ 29.6 & \citep{zhang2014four} \\
6 & 0.27 & 77.0 $\pm$ 14.0 & \citep{Simon:2004tf} \\
7 & 0.28 & 88.8 $\pm$ 36.6 & \citep{zhang2014four} \\
8 & 0.4 & 95.0 $\pm$ 17.0 & \citep{Simon:2004tf} \\
9 & 0.47 & 89.0 $\pm$ 50.0 & \citep{Ratsimbazafy:2017vga} \\
10 & 0.48 & 97.0 $\pm$ 62.0 & \citep{stern2010cosmic} \\
11 & 0.75 & 98.8 $\pm$ 33.6 & \citep{Borghi:2021rft} \\
12 & 0.88 & 90.0 $\pm$ 40.0 & \citep{stern2010cosmic} \\
13 & 0.9 & 117.0 $\pm$ 23.0 & \citep{Simon:2004tf} \\
14 & 1.3 & 168.0 $\pm$ 17.0 & \citep{Simon:2004tf} \\
15 & 1.43 & 177.0 $\pm$ 18.0 & \citep{Simon:2004tf} \\
16 & 1.53 & 140.0 $\pm$ 14.0 & \citep{Simon:2004tf} \\
17 & 1.75 & 202.0 $\pm$ 40.0 & \citep{Simon:2004tf} \\
18 & 0.1791* & 74.91 $\pm$ 5.57 & \citep{Moresco:2020fbm} \\
19 & 0.1993* & 74.96 $\pm$ 6.37 & \citep{Moresco:2020fbm} \\
20 & 0.3519* & 82.78 $\pm$ 14.65 & \citep{Moresco:2020fbm} \\
21 & 0.3802* & 83.0 $\pm$ 14.3 & \citep{Moresco:2020fbm} \\
22 & 0.4004* & 76.97 $\pm$ 11.12 & \citep{Moresco:2020fbm} \\
23 & 0.4247* & 87.08 $\pm$ 12.47 & \citep{Moresco:2020fbm} \\
24 & 0.4497* & 92.78 $\pm$ 14.07 & \citep{Moresco:2020fbm} \\
25 & 0.4783* & 80.91 $\pm$ 10.23 & \citep{Moresco:2020fbm} \\
26 & 0.5929* & 103.8 $\pm$ 14.0 & \citep{Moresco:2020fbm} \\
27 & 0.6797* & 91.6 $\pm$ 9.5 & \citep{Moresco:2020fbm} \\
28 & 0.7812* & 104.5 $\pm$ 13.3 & \citep{Moresco:2020fbm} \\
29 & 0.8754* & 125.1 $\pm$ 17.1 & \citep{Moresco:2020fbm} \\
30 & 1.037* & 153.7 $\pm$ 20.4 & \citep{Moresco:2020fbm} \\
31 & 1.363* & 160.0 $\pm$ 32.9 & \citep{Moresco:2020fbm} \\
32 & 1.965* & 186.5 $\pm$ 49.8 & \citep{Moresco:2020fbm} \\
\hline
\end{tabular}
\end{center}
\caption{
Observed values of the Hubble parameter with 1$\sigma$ uncertainties at 32 redshift points corresponding to cosmic chronometers (CC) observations. At the 15 redshift points with asterisks, the observations correlate with each pair of points i.e. covariances are present.
}
\label{table:H_CC_data}
\end{table}

We also consider the growth rate observations in our analysis. For the growth observations, we follow Avila et al. \citep{Avila:2022xad}. These observations consist of 11 uncorrelated $f$ data in the range $0.013 \leq z \leq 1.4$. The mean values and the standard deviations of the logarithmic growth rate $f$ are listed in Table~\ref{table:f_data}.

\begin{table*}
\begin{center}
\begin{tabular}{|c|c|c|c|c|c|}
\hline
Sr. & Survey & $z$ & $f$ & Refs. & Cosmological tracers \\
\hline
1 & ALFALFA & 0.013 & $0.56 \pm 0.07$ & \citep{Avila:2021dqv} & HI extragalactic sources \\
2 & 2dFGRS & 0.15 & $0.49 \pm 0.14$ & \citep{Hawkins:2002sg,Guzzo:2008ac} & galaxies \\
3 & GAMA & 0.18 & $0.49 \pm 0.12$ & \citep{Blake:2013nif} & multi-tracer: blue \& red gals. \\
4 & WiggleZ   & 0.22 & $0.60 \pm 0.10$ & \citep{Blake_2011} & galaxies \\
5 & SDSS    & 0.35 & $0.70 \pm 0.18$ & \citep{SDSS:2006lmn} & luminous red galaxies (LRG) \\
6 & GAMA   & 0.38 & $0.66 \pm 0.09$ & \citep{Blake:2013nif} & multi-tracer: blue \& red gals. \\
7 & WiggleZ & 0.41 & $0.70 \pm 0.07$ & \citep{Blake_2011} & galaxies \\
8 & 2SLAQ & 0.55 & $0.75 \pm 0.18$ & \citep{Ross:2006me} & LRG \& quasars \\
9 & WiggleZ & 0.60 & $0.73 \pm 0.07$ & \citep{Blake_2011} & galaxies \\
10 & VIMOS-VLT DS & 0.77 & $0.91 \pm 0.36$ & \citep{Guzzo:2008ac} & faint galaxies  \\
11 & 2QZ \& 2SLAQ & 1.40 & $0.90 \pm 0.24$ & \citep{daAngela:2006mf} & quasars \\
\hline
\end{tabular}
\end{center}
\caption{
Observed values of $f$ with 1$\sigma$ uncertainties at 11 redshift points corresponding to different surveys.
}
\label{table:f_data}
\end{table*}

\section{Methodology and results}
\label{sec-method_and_result}

From Eq.~\eqref{eq:final_sol_w}, we can see that computation of $w$ at a particular redshift $z$ depends on the values of $H$, $H'$, $f$, and $f'$ at the same redshift. Thus, in principle, computation of $w$ is possible from the data which are directly related to $H$, $H'$, $f$, and $f'$ and are at same redshift points. The cosmic chronometer data will be used to get $H$ and $H'$, whereas the logarithmic growth rate data will be utilized for finding the values of  $f$ and $f'$. However, note that, there is an issue that neither cosimc chronometer observations directly provide $H'$ nor the growth rate data directly provide $f'$. There is another issue that both these two kinds of data are not at the same redshift points. To overcome both these issues, we use the posterior Gaussian process regression (GPR) analysis. A brief discussion of the GPR is mentioned below.

\subsection{Brief overview of GPR}

In this study, we consider posterior GPR analysis \citep{williams1995gaussian,GpRasWil,Seikel_2012,Shafieloo_2012}. Consider a dataset consists of $n$ number of observational data points (either $n=32$ for cosmic chronometers observations or $n=11$ for growth rate observations). Two vectors $X$ and $Y$ represent the observation coordinates (here the observational redshift points) and mean values of a quantity (here observed values of either $H$ or $f$), respectively. Also, consider the data have observational errors through the covariance matrix $C$ i.e. $C = \text{Cov}[Y,Y]$. GPR analysis enables the prediction of mean values (represented by a vector $Y_*$) and covariances (represented by a matrix $\text{Cov}[Y_*,Y_*]$) for the same quantity at different target points $X_*$, using a kernel covariance function and a mean function. Here we assume the zero mean function to avoid any model dependence. The predicted values are obtained through the following expressions given as \citep{williams1995gaussian,GpRasWil,Seikel_2012,Shafieloo_2012}

\begin{eqnarray}
Y_* &=& K(X_*,X) \left[ K(X,X)+C \right]^{-1} Y,
\label{eq:GPR_mean_prediction} \\
\text{Cov}[Y_*,Y_*] &=& K(X_*,X_*) - K(X_*,X) \left[ K(X,X)+C \right]^{-1} K(X,X_*),
\label{eq:GPR_cov_prediction}
\end{eqnarray}

\noindent
where $K$ is the kernel matrix based on a specific kernel covariance function. Here we consider the squared-exponential kernel in which the covariance between two arbitrary points $x_1$ and $x_2$ are given as

\begin{equation}
k(x_1,x_2) = \sigma_f^2 e^{ -\frac{ (x_1-x_2)^2 }{2 l^2} },
\label{eq:kernel_SE}
\end{equation}

\noindent
where $k$ represents the matrix element of the main matrix $K$; $\sigma_f$ and $l$ are the corresponding kernel hyperparameters. We put optimal values of these hyperparameters for predictions in Eqs.~\eqref{eq:GPR_mean_prediction} and~\eqref{eq:GPR_cov_prediction}. These optimal values are determined by minimizing the negative log marginal likelihood ($\log P(Y|X)$) given as \citep{Seikel_2012}

\begin{equation}
\log P(Y|X) = -\frac{1}{2} Y^T \left[ K(X,X)+C \right]^{-1} Y -\frac{1}{2} \log |K(X,X)+C| -\frac{n}{2} \log{(2 \pi)},
\label{eq:log_marginal_likelihood}
\end{equation}

\noindent
where $|K(X,X)+C|$ is the determinant of the $K(X,X)+C$ matrix. The squared exponential (SE) function as its kernel is widely used due to its infinite differentiability, ensuring that the corresponding Gaussian process with this kernel exhibits smoothness, possessing mean-square derivatives of all orders \citep{GpRasWil}.

Additionally, GPR also predicts the gradient (first-order differentiation) of a quantity. The mean vector and covariance matrix corresponding to the first-order derivative are given by \citep{Seikel_2012}

\begin{eqnarray}
Y'_* &=& [K'(X,X_*)]^T \left[ K(X,X)+C \right]^{-1} Y,
\label{eq:derivative_mean_predictions} \\
\text{Cov}[Y'_*,Y'_*] &=& K''(X_*,X_*) - [K'(X,X_*)]^T \left[ K(X,X)+C \right]^{-1} K'(X,X_*),
\label{eq:derivative_cov_predictions}
\end{eqnarray}

\noindent
where prime and double prime denote the first and second order derivatives w.r.t the observational coordinate $x$ (here redshift $z$), respectively; $k'(x,x_*)$ and $k''(x_,x_*)$ are the partial derivatives of the kernel function given as

\begin{eqnarray}
k'(x,x_*) &=& \dfrac{\partial k(x,x_*)}{\partial x_*}, \\
k''(x_*,x_*) &=& \dfrac{\partial ^2 k(x_*,x_*)}{\partial x_* \partial x_*},
\label{eq:kernel_derivatives}
\end{eqnarray}

\noindent
respectively.

Similarly, the covariance matrix between the quantity and its first derivative is given as \citep{Seikel_2012}

\begin{equation}
\text{Cov}[Y_*,Y'_*] = K'(X_*,X_*) - [K(X,X_*)]^T \left[ K(X,X)+C \right]^{-1} K'(X,X_*),
\label{eq:cov_Y_Yp}
\end{equation}

\noindent
where $k'(x_*,x_*)$ is given as

\begin{equation}
k'(x_*,x_*) = \dfrac{\partial k(x_*,x_*)}{\partial x_*}.
\label{eq:kernel_derivatives_2}
\end{equation}

Thus, GPR predicts both the function and its derivative and also the covariances between them.

\subsection{Computation of $H$ and $H'$ and associated errors}

\begin{figure}
\centering
\includegraphics[width=0.45\textwidth]{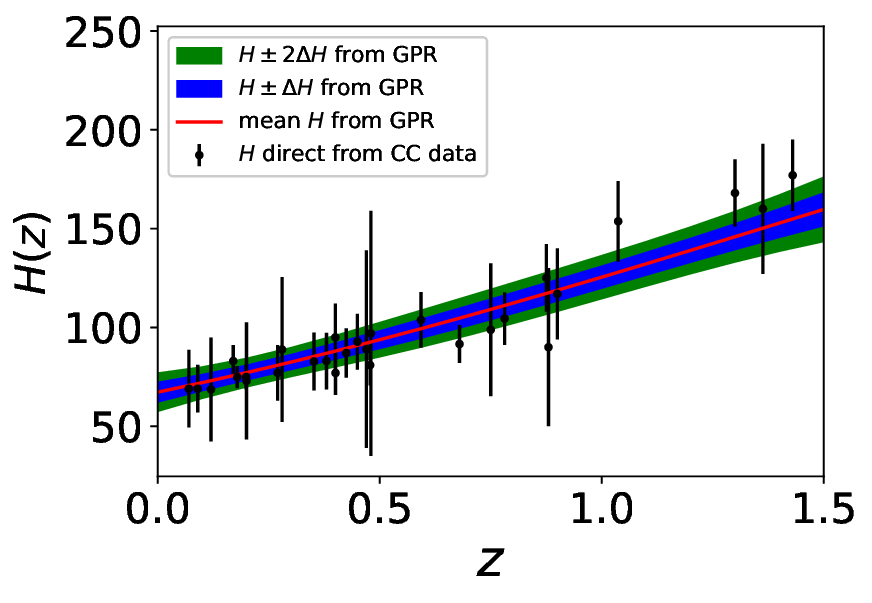} \\
\includegraphics[width=0.45\textwidth]{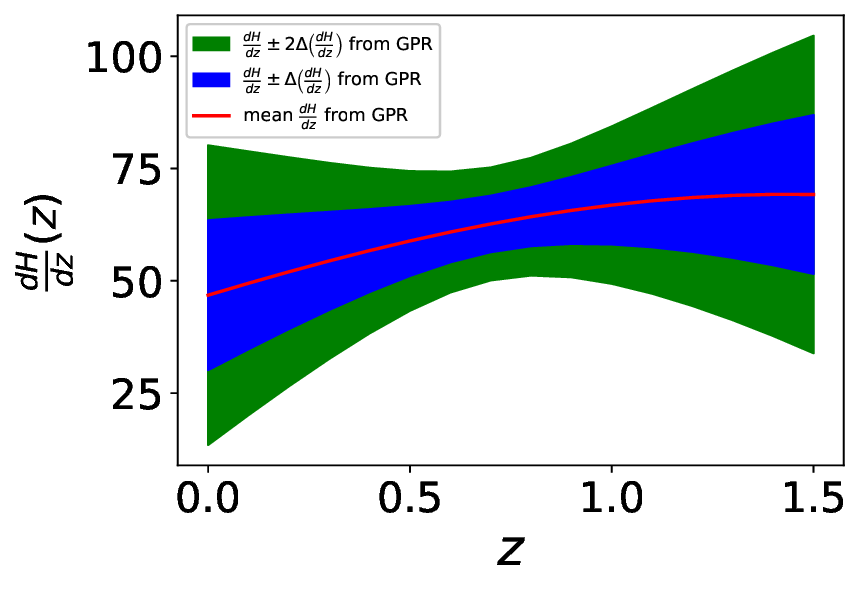}
\caption{
\label{fig:GPR_H_Hp}
GPR predictions for $H$ and $H'$ from cosmic chronometers (CC) data. The top and bottom panels correspond to the reconstruction of $H$ and $H'$ respectively. The red line, the blue region, and the green region correspond to the mean function, 1$\sigma$ and 2$\sigma$ confidence regions respectively. In the top panel, the black error bars correspond to the $H$ data directly obtained from the cosmic chronometer observations, mentioned in Table~\ref{table:H_CC_data}.
}
\end{figure}

We use GPR analysis to find $H$, $H'$, $\text{Var}[H]$, $\text{Var}[H']$, and $\text{Cov}[H,H']$ from the cosmic chronometers observations of Hubble parameter data at the target redshift points by the method mentioned in the previous subsection. The target redshift points should be such that it is compatible with the observed redshift points for both the cosmic chronometers and the growth rate observations. To have consistent target redshift points, we consider equally spaced redshift points in a redshift range as $0 \leq z \leq 1.5$ with the redshift interval $\Delta z=0.01$ so that we can get predictions for smooth functions. In Figure~\ref{fig:GPR_H_Hp}, we have shown the reconstructed functions for $H$ and $H'$ and the associated error from GPR analysis. The top and bottom panels correspond to the reconstruction of $H$ and $H'$ respectively. In both panels, the red line, the blue region, and the green region correspond to the mean function, 1$\sigma$, and 2$\sigma$ confidence regions for the estimation of the function respectively. In the top panel, the black error bars correspond to the $H$ data directly obtained from the cosmic chronometer observations. In this figure and the next figures, the notation $\Delta A$ (or $\Delta (A)$) represents the standard deviation of any quantity $A$ i.e. $\Delta A=\Delta (A)=\sqrt{\text{Var}[A]}$.

\subsection{Computation of $f$ and $f'$ and associated errors}

\begin{figure}
\centering
\includegraphics[width=0.45\textwidth]{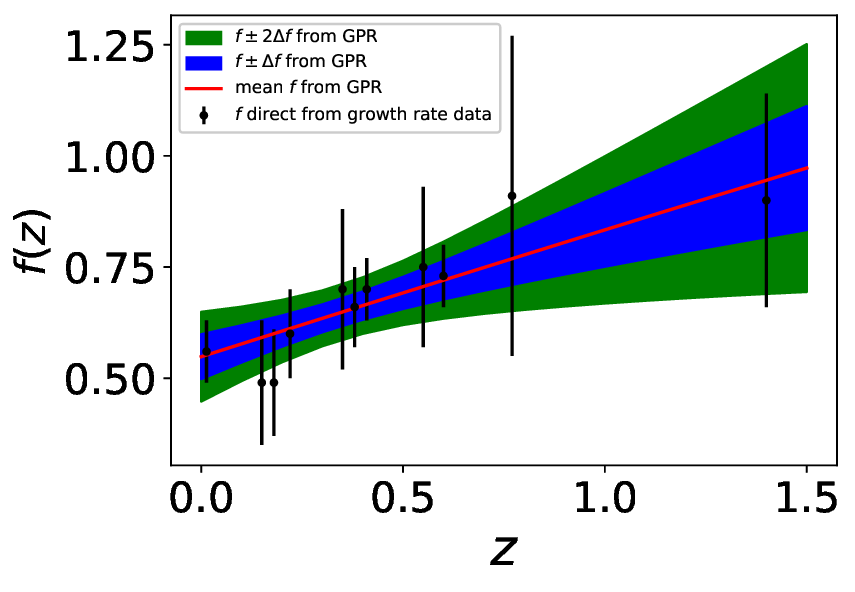} \\
\includegraphics[width=0.45\textwidth]{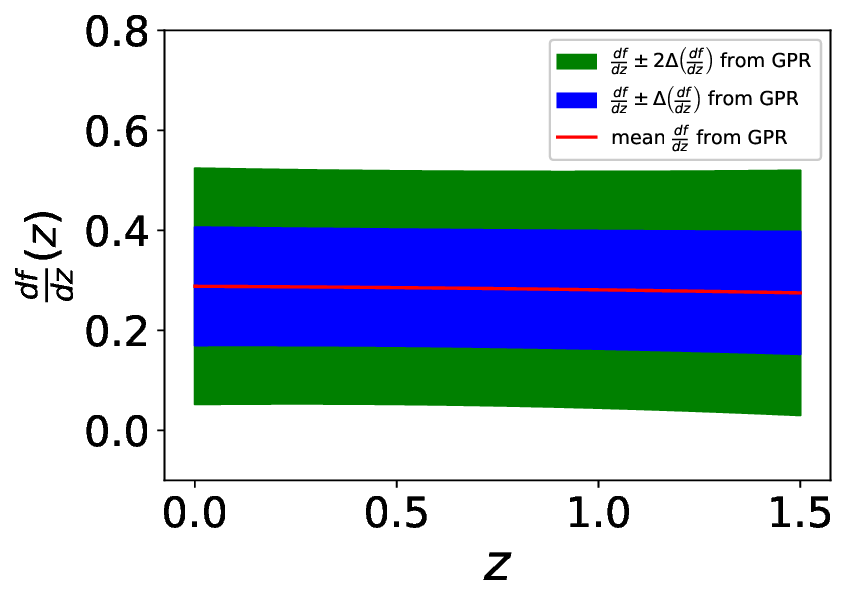}
\caption{
\label{fig:GPR_f_fp}
GPR predictions for $f$ and $f'$ from growth rate data. The top and bottom panels correspond to the reconstruction of $f$ and $f'$ respectively. The red line, the blue region, and the green region correspond to the mean function, 1$\sigma$ and 2$\sigma$ confidence regions respectively. In the top panel, the black error bars correspond to the $f$ data directly obtained from the growth rate observations, mentioned in Table~\ref{table:f_data}.
}
\end{figure}

Similarly we compute $f$, $f'$, $\text{Var}[f]$, $\text{Var}[f']$, and $\text{Cov}[f,f']$ from the growth rate data, mentioned in Table~\ref{table:f_data} using GPR. In Figure~\ref{fig:GPR_f_fp}, we have shown the reconstructed functions for $f$ and $f'$. The top and bottom panels correspond to the reconstructions of $f$ and $f'$ respectively. In both these panels, the red line, the blue region, and the green region correspond to the mean function, 1$\sigma$, and 2$\sigma$ confidence regions respectively. In the top panel, we also plot the $f$ data obtained directly from the growth rate observations by black error bars.

\subsection{Computation of $w$ and associated errors}

\begin{figure}
\centering
\includegraphics[width=0.45\textwidth]{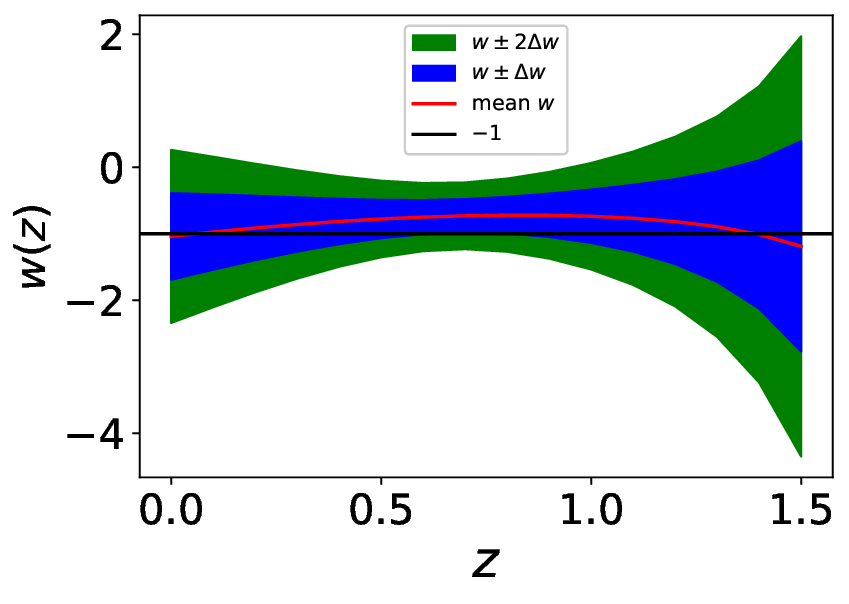}
\caption{
\label{fig:wde}
Redshift evolution of dark energy equation of state. The red line, the blue region, and the green region correspond to the mean function, 1$\sigma$ and 2$\sigma$ confidence regions respectively. The black line represents a particular value $w=-1$ which corresponds to the popular $\Lambda$CDM model.
}
\end{figure}

Now, we have reconstructed values of $H$, $H'$, $f$, and $f'$ at each redshift point. We put these values in Eq.~\eqref{eq:final_sol_w} to compute $w$ at the same redshift points. In this way, we reconstruct a mean function for $w$. Similarly, we compute the standard deviations in the estimation of $w$ using Eq.~\eqref{eq:var_w}. In Figure~\ref{fig:wde}, we have shown the reconstructed function of $w$. The red line, the blue region, and the green region correspond to the mean function, 1$\sigma$, and 2$\sigma$ confidence regions respectively.

In Figure~\ref{fig:wde}, we have also plotted a black line representing $w=-1$ to compare the results with the standard $\Lambda$CDM model. We find that the 1$\sigma$ or 2$\sigma$ confidence regions are smaller in the range $0.5 \lesssim z \lesssim 1$ i.e. in this redshift range the estimation of $w$ is comparatively tightly constrained. We find that the $\Lambda$CDM model is situated almost at the lower border of the reconstructed 1$\sigma$ region at a very short redshift range, $0.6 \lesssim z \lesssim 0.85$. Except at this short redshift range, in other regions, the $\Lambda$CDM model is within the 1$\sigma$ region. Note that, it is well within the 2$\sigma$ confidence region for the full redshift range, considered in this analysis.

In Figure~\ref{fig:wde}, we can see that the large (more than half) portions of the blue (1$\sigma$) and the green (2$\sigma$) regions are above the black line ($\Lambda$CDM) i.e. at the non-phantom region ($w>-1$). So, non-phantom values of $w$ are marginally favored over the phantom values.

In Figure~\ref{fig:wde}, we can see that the reconstructed mean function of $w$ (the red line) is not a straight line. This suggests a dynamical behavior of the equation of state of dark energy. However, the evidence of dynamical dark energy is not very significant because one can easily draw some straight lines corresponding to constant values of $w$ well within the 1$\sigma$ (blue) region. We are not explicitly showing these straight lines to keep the figure simple but it is obvious from the 1$\sigma$ region.

In summary, a non-parametric model-independent study suggests that there is no strong evidence to go beyond the $\Lambda$CDM model corresponding to cosmic chronometer and growth observations. In the future, this study can be further extended to include other cosmological observations to study the dark energy evolution in more detail using similar approaches.

\section{Conclusion}
\label{sec-conclusion}

We explore the behavior of dark energy without considering any cosmological model or parameterization. We only consider three assumptions: (I) the background dynamics is described by the flat FLRW metric, (II) no interaction between dark energy and matter sectors is considered, and (III) we restricted ourselve on such cosmic scales where sub-Horizon approximation and linear perturbations are applicable for the growth of matter inhomogeneity.sss

This is the first study to combine both background and perturbation observations to reconstruct the dynamics of dark energy. We consider cosmic chronometer data for the Hubble parameter and growth rate observations for the growth rate $f$ simultaneously. Using posterior Gaussian process regression (GPR) analysis, we reconstruct the smooth functions of the Hubble parameter $H$, its derivative $H'$, and the associated errors from the cosmic chronometer data. Similarly, from the growth rate data, we compute $f$, $f'$, and the associated errors using GPR analysis. We use these values for a robust reconstruction of the dark energy equation of state parameter and the associated errors using a unique equation derived from the flat FLRW background geometry and the Newtonian perturbation theory on linear scales.

We observe that the errors in the reconstruction of $w$ are narrower in the redshift range $0.5 \lesssim z \lesssim 1$ i.e. the estimation of $w$ is more tightly constrained within this redshift interval.

We compare the reconstructed $w$ with the standard $\Lambda$CDM model and we find that the $\Lambda$CDM model is at almost the lower boundary of the 1$\sigma$ reconstructed region at redshift interval $0.6 \lesssim z \lesssim 0.85$. In other redshift regions of study, it is well within the 1$\sigma$ region.

We observe a preference for regions where dark energy exhibits non-phantom behavior ($w > -1$).

We also find some indications of the dynamical evolution of dark energy, although the evidence is rather insignificant.

Although this is the first truly model-independent non-parametric reconstruction of the dark energy equation of state parameter $w$, the obvious weak point of this is the use of linear perturbations only. Future work on this using non-linear perturbation equations will hold the key for the determination of the final fate of the $\Lambda$CDM model. This study can also be extended to include other key cosmological observations and we leave this for future study.

In summary, we find no strong evidence to go beyond the standard $\Lambda$CDM model, when we consider a non-parametric model-independent study of dark energy from cosmic chronometer and growth rate observation. Although the results obtained here are well established, the methodology used in this analysis is unique in the sense it is completely model independent and nonparametric. So, in this direction, more detailed analyses for the generalization of this study with similar kind of methodologies will be taken up in future studies.

\acknowledgments
BRD would like to acknowledge IISER Kolkata for the financial support through the postdoctoral fellowship. BRD was also partially suported by the South African Radio Astronomy Observatory and  National Research Foundation (Grant No. 75415).

\bibliographystyle{apsrev4-1}
\bibliography{refsgpde}

\begin{thebibliography}{125}%
\makeatletter
\providecommand \@ifxundefined [1]{%
 \@ifx{#1\undefined}
}%
\providecommand \@ifnum [1]{%
 \ifnum #1\expandafter \@firstoftwo
 \else \expandafter \@secondoftwo
 \fi
}%
\providecommand \@ifx [1]{%
 \ifx #1\expandafter \@firstoftwo
 \else \expandafter \@secondoftwo
 \fi
}%
\providecommand \natexlab [1]{#1}%
\providecommand \enquote  [1]{``#1''}%
\providecommand \bibnamefont  [1]{#1}%
\providecommand \bibfnamefont [1]{#1}%
\providecommand \citenamefont [1]{#1}%
\providecommand \href@noop [0]{\@secondoftwo}%
\providecommand \href [0]{\begingroup \@sanitize@url \@href}%
\providecommand \@href[1]{\@@startlink{#1}\@@href}%
\providecommand \@@href[1]{\endgroup#1\@@endlink}%
\providecommand \@sanitize@url [0]{\catcode `\\12\catcode `\$12\catcode
  `\&12\catcode `\#12\catcode `\^12\catcode `\_12\catcode `\%12\relax}%
\providecommand \@@startlink[1]{}%
\providecommand \@@endlink[0]{}%
\providecommand \url  [0]{\begingroup\@sanitize@url \@url }%
\providecommand \@url [1]{\endgroup\@href {#1}{\urlprefix }}%
\providecommand \urlprefix  [0]{URL }%
\providecommand \Eprint [0]{\href }%
\providecommand \doibase [0]{http://dx.doi.org/}%
\providecommand \selectlanguage [0]{\@gobble}%
\providecommand \bibinfo  [0]{\@secondoftwo}%
\providecommand \bibfield  [0]{\@secondoftwo}%
\providecommand \translation [1]{[#1]}%
\providecommand \BibitemOpen [0]{}%
\providecommand \bibitemStop [0]{}%
\providecommand \bibitemNoStop [0]{.\EOS\space}%
\providecommand \EOS [0]{\spacefactor3000\relax}%
\providecommand \BibitemShut  [1]{\csname bibitem#1\endcsname}%
\let\auto@bib@innerbib\@empty
\bibitem [{\citenamefont {Perlmutter}\ \emph {et~al.}(1998)\citenamefont
  {Perlmutter} \emph {et~al.}}]{SupernovaCosmologyProject:1997zqe}%
  \BibitemOpen
  \bibfield  {author} {\bibinfo {author} {\bibfnamefont {S.}~\bibnamefont
  {Perlmutter}} \emph {et~al.} (\bibinfo {collaboration} {Supernova Cosmology
  Project}),\ }\href {\doibase 10.1038/34124} {\bibfield  {journal} {\bibinfo
  {journal} {Nature}\ }\textbf {\bibinfo {volume} {391}},\ \bibinfo {pages}
  {51} (\bibinfo {year} {1998})},\ \Eprint
  {http://arxiv.org/abs/astro-ph/9712212} {arXiv:astro-ph/9712212} \BibitemShut
  {NoStop}%
\bibitem [{\citenamefont {Riess}\ \emph {et~al.}(1998)\citenamefont {Riess}
  \emph {et~al.}}]{SupernovaSearchTeam:1998fmf}%
  \BibitemOpen
  \bibfield  {author} {\bibinfo {author} {\bibfnamefont {A.~G.}\ \bibnamefont
  {Riess}} \emph {et~al.} (\bibinfo {collaboration} {Supernova Search Team}),\
  }\href {\doibase 10.1086/300499} {\bibfield  {journal} {\bibinfo  {journal}
  {Astron. J.}\ }\textbf {\bibinfo {volume} {116}},\ \bibinfo {pages} {1009}
  (\bibinfo {year} {1998})},\ \Eprint {http://arxiv.org/abs/astro-ph/9805201}
  {arXiv:astro-ph/9805201} \BibitemShut {NoStop}%
\bibitem [{\citenamefont {Perlmutter}\ \emph {et~al.}(1999)\citenamefont
  {Perlmutter} \emph {et~al.}}]{SupernovaCosmologyProject:1998vns}%
  \BibitemOpen
  \bibfield  {author} {\bibinfo {author} {\bibfnamefont {S.}~\bibnamefont
  {Perlmutter}} \emph {et~al.} (\bibinfo {collaboration} {Supernova Cosmology
  Project}),\ }\href {\doibase 10.1086/307221} {\bibfield  {journal} {\bibinfo
  {journal} {Astrophys. J.}\ }\textbf {\bibinfo {volume} {517}},\ \bibinfo
  {pages} {565} (\bibinfo {year} {1999})},\ \Eprint
  {http://arxiv.org/abs/astro-ph/9812133} {arXiv:astro-ph/9812133} \BibitemShut
  {NoStop}%
\bibitem [{\citenamefont {{Wright}}(2011)}]{2011NatPh...7Q.833W}%
  \BibitemOpen
  \bibfield  {author} {\bibinfo {author} {\bibfnamefont {A.}~\bibnamefont
  {{Wright}}},\ }\href {\doibase 10.1038/nphys2131} {\bibfield  {journal}
  {\bibinfo  {journal} {Nature Physics}\ }\textbf {\bibinfo {volume} {7}},\
  \bibinfo {pages} {833} (\bibinfo {year} {2011})}\BibitemShut {NoStop}%
\bibitem [{\citenamefont {Ade}\ \emph {et~al.}(2014)\citenamefont {Ade} \emph
  {et~al.}}]{Planck:2013pxb}%
  \BibitemOpen
  \bibfield  {author} {\bibinfo {author} {\bibfnamefont {P.~A.~R.}\
  \bibnamefont {Ade}} \emph {et~al.} (\bibinfo {collaboration} {Planck}),\
  }\href {\doibase 10.1051/0004-6361/201321591} {\bibfield  {journal} {\bibinfo
   {journal} {Astron. Astrophys.}\ }\textbf {\bibinfo {volume} {571}},\
  \bibinfo {pages} {A16} (\bibinfo {year} {2014})},\ \Eprint
  {http://arxiv.org/abs/1303.5076} {arXiv:1303.5076 [astro-ph.CO]} \BibitemShut
  {NoStop}%
\bibitem [{\citenamefont {Ade}\ \emph {et~al.}(2016)\citenamefont {Ade} \emph
  {et~al.}}]{Planck:2015fie}%
  \BibitemOpen
  \bibfield  {author} {\bibinfo {author} {\bibfnamefont {P.~A.~R.}\
  \bibnamefont {Ade}} \emph {et~al.} (\bibinfo {collaboration} {Planck}),\
  }\href {\doibase 10.1051/0004-6361/201525830} {\bibfield  {journal} {\bibinfo
   {journal} {Astron. Astrophys.}\ }\textbf {\bibinfo {volume} {594}},\
  \bibinfo {pages} {A13} (\bibinfo {year} {2016})},\ \Eprint
  {http://arxiv.org/abs/1502.01589} {arXiv:1502.01589 [astro-ph.CO]}
  \BibitemShut {NoStop}%
\bibitem [{\citenamefont {Aghanim}\ \emph {et~al.}(2020)\citenamefont {Aghanim}
  \emph {et~al.}}]{Planck:2018vyg}%
  \BibitemOpen
  \bibfield  {author} {\bibinfo {author} {\bibfnamefont {N.}~\bibnamefont
  {Aghanim}} \emph {et~al.} (\bibinfo {collaboration} {Planck}),\ }\href
  {\doibase 10.1051/0004-6361/201833910} {\bibfield  {journal} {\bibinfo
  {journal} {Astron. Astrophys.}\ }\textbf {\bibinfo {volume} {641}},\ \bibinfo
  {pages} {A6} (\bibinfo {year} {2020})},\ \bibinfo {note} {[Erratum:
  Astron.Astrophys. 652, C4 (2021)]},\ \Eprint
  {http://arxiv.org/abs/1807.06209} {arXiv:1807.06209 [astro-ph.CO]}
  \BibitemShut {NoStop}%
\bibitem [{\citenamefont {Alam}\ \emph {et~al.}(2017)\citenamefont {Alam} \emph
  {et~al.}}]{BOSS:2016wmc}%
  \BibitemOpen
  \bibfield  {author} {\bibinfo {author} {\bibfnamefont {S.}~\bibnamefont
  {Alam}} \emph {et~al.} (\bibinfo {collaboration} {BOSS}),\ }\href {\doibase
  10.1093/mnras/stx721} {\bibfield  {journal} {\bibinfo  {journal} {Mon. Not.
  Roy. Astron. Soc.}\ }\textbf {\bibinfo {volume} {470}},\ \bibinfo {pages}
  {2617} (\bibinfo {year} {2017})},\ \Eprint {http://arxiv.org/abs/1607.03155}
  {arXiv:1607.03155 [astro-ph.CO]} \BibitemShut {NoStop}%
\bibitem [{\citenamefont {Alam}\ \emph {et~al.}(2021)\citenamefont {Alam} \emph
  {et~al.}}]{eBOSS:2020yzd}%
  \BibitemOpen
  \bibfield  {author} {\bibinfo {author} {\bibfnamefont {S.}~\bibnamefont
  {Alam}} \emph {et~al.} (\bibinfo {collaboration} {eBOSS}),\ }\href {\doibase
  10.1103/PhysRevD.103.083533} {\bibfield  {journal} {\bibinfo  {journal}
  {Phys. Rev. D}\ }\textbf {\bibinfo {volume} {103}},\ \bibinfo {pages}
  {083533} (\bibinfo {year} {2021})},\ \Eprint
  {http://arxiv.org/abs/2007.08991} {arXiv:2007.08991 [astro-ph.CO]}
  \BibitemShut {NoStop}%
\bibitem [{\citenamefont {Hou}\ \emph {et~al.}(2020)\citenamefont {Hou} \emph
  {et~al.}}]{Hou:2020rse}%
  \BibitemOpen
  \bibfield  {author} {\bibinfo {author} {\bibfnamefont {J.}~\bibnamefont
  {Hou}} \emph {et~al.},\ }\href {\doibase 10.1093/mnras/staa3234} {\bibfield
  {journal} {\bibinfo  {journal} {Mon. Not. Roy. Astron. Soc.}\ }\textbf
  {\bibinfo {volume} {500}},\ \bibinfo {pages} {1201} (\bibinfo {year}
  {2020})},\ \Eprint {http://arxiv.org/abs/2007.08998} {arXiv:2007.08998
  [astro-ph.CO]} \BibitemShut {NoStop}%
\bibitem [{\citenamefont {Jimenez}\ and\ \citenamefont
  {Loeb}(2002)}]{Jimenez:2001gg}%
  \BibitemOpen
  \bibfield  {author} {\bibinfo {author} {\bibfnamefont {R.}~\bibnamefont
  {Jimenez}}\ and\ \bibinfo {author} {\bibfnamefont {A.}~\bibnamefont {Loeb}},\
  }\href {\doibase 10.1086/340549} {\bibfield  {journal} {\bibinfo  {journal}
  {Astrophys. J.}\ }\textbf {\bibinfo {volume} {573}},\ \bibinfo {pages} {37}
  (\bibinfo {year} {2002})},\ \Eprint {http://arxiv.org/abs/astro-ph/0106145}
  {arXiv:astro-ph/0106145} \BibitemShut {NoStop}%
\bibitem [{\citenamefont {Pinho}\ \emph {et~al.}(2018)\citenamefont {Pinho},
  \citenamefont {Casas},\ and\ \citenamefont {Amendola}}]{Pinho:2018unz}%
  \BibitemOpen
  \bibfield  {author} {\bibinfo {author} {\bibfnamefont {A.~M.}\ \bibnamefont
  {Pinho}}, \bibinfo {author} {\bibfnamefont {S.}~\bibnamefont {Casas}}, \ and\
  \bibinfo {author} {\bibfnamefont {L.}~\bibnamefont {Amendola}},\ }\href
  {\doibase 10.1088/1475-7516/2018/11/027} {\bibfield  {journal} {\bibinfo
  {journal} {JCAP}\ }\textbf {\bibinfo {volume} {11}},\ \bibinfo {pages} {027}
  (\bibinfo {year} {2018})},\ \Eprint {http://arxiv.org/abs/1805.00027}
  {arXiv:1805.00027 [astro-ph.CO]} \BibitemShut {NoStop}%
\bibitem [{\citenamefont {Cao}\ and\ \citenamefont
  {Ratra}(2023)}]{Cao:2023eja}%
  \BibitemOpen
  \bibfield  {author} {\bibinfo {author} {\bibfnamefont {S.}~\bibnamefont
  {Cao}}\ and\ \bibinfo {author} {\bibfnamefont {B.}~\bibnamefont {Ratra}},\
  }\href {\doibase 10.1103/PhysRevD.107.103521} {\bibfield  {journal} {\bibinfo
   {journal} {Phys. Rev. D}\ }\textbf {\bibinfo {volume} {107}},\ \bibinfo
  {pages} {103521} (\bibinfo {year} {2023})},\ \Eprint
  {http://arxiv.org/abs/2302.14203} {arXiv:2302.14203 [astro-ph.CO]}
  \BibitemShut {NoStop}%
\bibitem [{\citenamefont {Peebles}\ and\ \citenamefont
  {Ratra}(2003)}]{Peebles:2002gy}%
  \BibitemOpen
  \bibfield  {author} {\bibinfo {author} {\bibfnamefont {P.~J.~E.}\
  \bibnamefont {Peebles}}\ and\ \bibinfo {author} {\bibfnamefont
  {B.}~\bibnamefont {Ratra}},\ }\href {\doibase 10.1103/RevModPhys.75.559}
  {\bibfield  {journal} {\bibinfo  {journal} {Rev. Mod. Phys.}\ }\textbf
  {\bibinfo {volume} {75}},\ \bibinfo {pages} {559} (\bibinfo {year} {2003})},\
  \Eprint {http://arxiv.org/abs/astro-ph/0207347} {arXiv:astro-ph/0207347}
  \BibitemShut {NoStop}%
\bibitem [{\citenamefont {Copeland}\ \emph {et~al.}(2006)\citenamefont
  {Copeland}, \citenamefont {Sami},\ and\ \citenamefont
  {Tsujikawa}}]{Copeland:2006wr}%
  \BibitemOpen
  \bibfield  {author} {\bibinfo {author} {\bibfnamefont {E.~J.}\ \bibnamefont
  {Copeland}}, \bibinfo {author} {\bibfnamefont {M.}~\bibnamefont {Sami}}, \
  and\ \bibinfo {author} {\bibfnamefont {S.}~\bibnamefont {Tsujikawa}},\ }\href
  {\doibase 10.1142/S021827180600942X} {\bibfield  {journal} {\bibinfo
  {journal} {Int. J. Mod. Phys. D}\ }\textbf {\bibinfo {volume} {15}},\
  \bibinfo {pages} {1753} (\bibinfo {year} {2006})},\ \Eprint
  {http://arxiv.org/abs/hep-th/0603057} {arXiv:hep-th/0603057} \BibitemShut
  {NoStop}%
\bibitem [{\citenamefont {Yoo}\ and\ \citenamefont
  {Watanabe}(2012)}]{Yoo:2012ug}%
  \BibitemOpen
  \bibfield  {author} {\bibinfo {author} {\bibfnamefont {J.}~\bibnamefont
  {Yoo}}\ and\ \bibinfo {author} {\bibfnamefont {Y.}~\bibnamefont {Watanabe}},\
  }\href {\doibase 10.1142/S0218271812300029} {\bibfield  {journal} {\bibinfo
  {journal} {Int. J. Mod. Phys. D}\ }\textbf {\bibinfo {volume} {21}},\
  \bibinfo {pages} {1230002} (\bibinfo {year} {2012})},\ \Eprint
  {http://arxiv.org/abs/1212.4726} {arXiv:1212.4726 [astro-ph.CO]} \BibitemShut
  {NoStop}%
\bibitem [{\citenamefont {Lonappan}\ \emph {et~al.}(2018)\citenamefont
  {Lonappan}, \citenamefont {Kumar}, \citenamefont {Ruchika}, \citenamefont
  {Dinda},\ and\ \citenamefont {Sen}}]{Lonappan:2017lzt}%
  \BibitemOpen
  \bibfield  {author} {\bibinfo {author} {\bibfnamefont {A.~I.}\ \bibnamefont
  {Lonappan}}, \bibinfo {author} {\bibfnamefont {S.}~\bibnamefont {Kumar}},
  \bibinfo {author} {\bibnamefont {Ruchika}}, \bibinfo {author} {\bibfnamefont
  {B.~R.}\ \bibnamefont {Dinda}}, \ and\ \bibinfo {author} {\bibfnamefont
  {A.~A.}\ \bibnamefont {Sen}},\ }\href {\doibase 10.1103/PhysRevD.97.043524}
  {\bibfield  {journal} {\bibinfo  {journal} {Phys. Rev. D}\ }\textbf {\bibinfo
  {volume} {97}},\ \bibinfo {pages} {043524} (\bibinfo {year} {2018})},\
  \Eprint {http://arxiv.org/abs/1707.00603} {arXiv:1707.00603 [astro-ph.CO]}
  \BibitemShut {NoStop}%
\bibitem [{\citenamefont {Dinda}(2017)}]{Dinda:2017swh}%
  \BibitemOpen
  \bibfield  {author} {\bibinfo {author} {\bibfnamefont {B.~R.}\ \bibnamefont
  {Dinda}},\ }\href {\doibase 10.1088/1475-7516/2017/09/035} {\bibfield
  {journal} {\bibinfo  {journal} {JCAP}\ }\textbf {\bibinfo {volume} {09}},\
  \bibinfo {pages} {035} (\bibinfo {year} {2017})},\ \Eprint
  {http://arxiv.org/abs/1705.00657} {arXiv:1705.00657 [astro-ph.CO]}
  \BibitemShut {NoStop}%
\bibitem [{\citenamefont {Dinda}\ \emph
  {et~al.}(2018{\natexlab{a}})\citenamefont {Dinda}, \citenamefont {Sen},\ and\
  \citenamefont {Choudhury}}]{Dinda:2018uwm}%
  \BibitemOpen
  \bibfield  {author} {\bibinfo {author} {\bibfnamefont {B.~R.}\ \bibnamefont
  {Dinda}}, \bibinfo {author} {\bibfnamefont {A.~A.}\ \bibnamefont {Sen}}, \
  and\ \bibinfo {author} {\bibfnamefont {T.~R.}\ \bibnamefont {Choudhury}},\
  }\href@noop {} {\  (\bibinfo {year} {2018}{\natexlab{a}})},\ \Eprint
  {http://arxiv.org/abs/1804.11137} {arXiv:1804.11137 [astro-ph.CO]}
  \BibitemShut {NoStop}%
\bibitem [{\citenamefont {Bamba}\ \emph {et~al.}(2012)\citenamefont {Bamba},
  \citenamefont {Capozziello}, \citenamefont {Nojiri},\ and\ \citenamefont
  {Odintsov}}]{Bamba:2012cp}%
  \BibitemOpen
  \bibfield  {author} {\bibinfo {author} {\bibfnamefont {K.}~\bibnamefont
  {Bamba}}, \bibinfo {author} {\bibfnamefont {S.}~\bibnamefont {Capozziello}},
  \bibinfo {author} {\bibfnamefont {S.}~\bibnamefont {Nojiri}}, \ and\ \bibinfo
  {author} {\bibfnamefont {S.~D.}\ \bibnamefont {Odintsov}},\ }\href {\doibase
  10.1007/s10509-012-1181-8} {\bibfield  {journal} {\bibinfo  {journal}
  {Astrophys. Space Sci.}\ }\textbf {\bibinfo {volume} {342}},\ \bibinfo
  {pages} {155} (\bibinfo {year} {2012})},\ \Eprint
  {http://arxiv.org/abs/1205.3421} {arXiv:1205.3421 [gr-qc]} \BibitemShut
  {NoStop}%
\bibitem [{\citenamefont {Clifton}\ \emph {et~al.}(2012)\citenamefont
  {Clifton}, \citenamefont {Ferreira}, \citenamefont {Padilla},\ and\
  \citenamefont {Skordis}}]{Clifton:2011jh}%
  \BibitemOpen
  \bibfield  {author} {\bibinfo {author} {\bibfnamefont {T.}~\bibnamefont
  {Clifton}}, \bibinfo {author} {\bibfnamefont {P.~G.}\ \bibnamefont
  {Ferreira}}, \bibinfo {author} {\bibfnamefont {A.}~\bibnamefont {Padilla}}, \
  and\ \bibinfo {author} {\bibfnamefont {C.}~\bibnamefont {Skordis}},\ }\href
  {\doibase 10.1016/j.physrep.2012.01.001} {\bibfield  {journal} {\bibinfo
  {journal} {Phys. Rept.}\ }\textbf {\bibinfo {volume} {513}},\ \bibinfo
  {pages} {1} (\bibinfo {year} {2012})},\ \Eprint
  {http://arxiv.org/abs/1106.2476} {arXiv:1106.2476 [astro-ph.CO]} \BibitemShut
  {NoStop}%
\bibitem [{\citenamefont {Koyama}(2016)}]{Koyama:2015vza}%
  \BibitemOpen
  \bibfield  {author} {\bibinfo {author} {\bibfnamefont {K.}~\bibnamefont
  {Koyama}},\ }\href {\doibase 10.1088/0034-4885/79/4/046902} {\bibfield
  {journal} {\bibinfo  {journal} {Rept. Prog. Phys.}\ }\textbf {\bibinfo
  {volume} {79}},\ \bibinfo {pages} {046902} (\bibinfo {year} {2016})},\
  \Eprint {http://arxiv.org/abs/1504.04623} {arXiv:1504.04623 [astro-ph.CO]}
  \BibitemShut {NoStop}%
\bibitem [{\citenamefont {Tsujikawa}(2010)}]{Tsujikawa:2010zza}%
  \BibitemOpen
  \bibfield  {author} {\bibinfo {author} {\bibfnamefont {S.}~\bibnamefont
  {Tsujikawa}},\ }\href {\doibase 10.1007/978-3-642-10598-2_3} {\bibfield
  {journal} {\bibinfo  {journal} {Lect. Notes Phys.}\ }\textbf {\bibinfo
  {volume} {800}},\ \bibinfo {pages} {99} (\bibinfo {year} {2010})},\ \Eprint
  {http://arxiv.org/abs/1101.0191} {arXiv:1101.0191 [gr-qc]} \BibitemShut
  {NoStop}%
\bibitem [{\citenamefont {Joyce}\ \emph {et~al.}(2016)\citenamefont {Joyce},
  \citenamefont {Lombriser},\ and\ \citenamefont {Schmidt}}]{Joyce:2016vqv}%
  \BibitemOpen
  \bibfield  {author} {\bibinfo {author} {\bibfnamefont {A.}~\bibnamefont
  {Joyce}}, \bibinfo {author} {\bibfnamefont {L.}~\bibnamefont {Lombriser}}, \
  and\ \bibinfo {author} {\bibfnamefont {F.}~\bibnamefont {Schmidt}},\ }\href
  {\doibase 10.1146/annurev-nucl-102115-044553} {\bibfield  {journal} {\bibinfo
   {journal} {Ann. Rev. Nucl. Part. Sci.}\ }\textbf {\bibinfo {volume} {66}},\
  \bibinfo {pages} {95} (\bibinfo {year} {2016})},\ \Eprint
  {http://arxiv.org/abs/1601.06133} {arXiv:1601.06133 [astro-ph.CO]}
  \BibitemShut {NoStop}%
\bibitem [{\citenamefont {Dinda}\ \emph
  {et~al.}(2018{\natexlab{b}})\citenamefont {Dinda}, \citenamefont
  {Wali~Hossain},\ and\ \citenamefont {Sen}}]{Dinda:2017lpz}%
  \BibitemOpen
  \bibfield  {author} {\bibinfo {author} {\bibfnamefont {B.~R.}\ \bibnamefont
  {Dinda}}, \bibinfo {author} {\bibfnamefont {M.}~\bibnamefont {Wali~Hossain}},
  \ and\ \bibinfo {author} {\bibfnamefont {A.~A.}\ \bibnamefont {Sen}},\ }\href
  {\doibase 10.1088/1475-7516/2018/01/045} {\bibfield  {journal} {\bibinfo
  {journal} {JCAP}\ }\textbf {\bibinfo {volume} {01}},\ \bibinfo {pages} {045}
  (\bibinfo {year} {2018}{\natexlab{b}})},\ \Eprint
  {http://arxiv.org/abs/1706.00567} {arXiv:1706.00567 [astro-ph.CO]}
  \BibitemShut {NoStop}%
\bibitem [{\citenamefont {Dinda}(2018)}]{Dinda:2018eyt}%
  \BibitemOpen
  \bibfield  {author} {\bibinfo {author} {\bibfnamefont {B.~R.}\ \bibnamefont
  {Dinda}},\ }\href {\doibase 10.1088/1475-7516/2018/06/017} {\bibfield
  {journal} {\bibinfo  {journal} {JCAP}\ }\textbf {\bibinfo {volume} {06}},\
  \bibinfo {pages} {017} (\bibinfo {year} {2018})},\ \Eprint
  {http://arxiv.org/abs/1801.01741} {arXiv:1801.01741 [astro-ph.CO]}
  \BibitemShut {NoStop}%
\bibitem [{\citenamefont {Zhang}\ \emph {et~al.}(2020)\citenamefont {Zhang},
  \citenamefont {Dinda}, \citenamefont {Hossain}, \citenamefont {Sen},\ and\
  \citenamefont {Luo}}]{Zhang:2020qkd}%
  \BibitemOpen
  \bibfield  {author} {\bibinfo {author} {\bibfnamefont {J.}~\bibnamefont
  {Zhang}}, \bibinfo {author} {\bibfnamefont {B.~R.}\ \bibnamefont {Dinda}},
  \bibinfo {author} {\bibfnamefont {M.~W.}\ \bibnamefont {Hossain}}, \bibinfo
  {author} {\bibfnamefont {A.~A.}\ \bibnamefont {Sen}}, \ and\ \bibinfo
  {author} {\bibfnamefont {W.}~\bibnamefont {Luo}},\ }\href {\doibase
  10.1103/PhysRevD.102.043510} {\bibfield  {journal} {\bibinfo  {journal}
  {Phys. Rev. D}\ }\textbf {\bibinfo {volume} {102}},\ \bibinfo {pages}
  {043510} (\bibinfo {year} {2020})},\ \Eprint
  {http://arxiv.org/abs/2004.12659} {arXiv:2004.12659 [astro-ph.CO]}
  \BibitemShut {NoStop}%
\bibitem [{\citenamefont {Dinda}\ \emph {et~al.}(2023)\citenamefont {Dinda},
  \citenamefont {Hossain},\ and\ \citenamefont {Sen}}]{Dinda:2022ixi}%
  \BibitemOpen
  \bibfield  {author} {\bibinfo {author} {\bibfnamefont {B.~R.}\ \bibnamefont
  {Dinda}}, \bibinfo {author} {\bibfnamefont {M.~W.}\ \bibnamefont {Hossain}},
  \ and\ \bibinfo {author} {\bibfnamefont {A.~A.}\ \bibnamefont {Sen}},\ }\href
  {\doibase 10.1007/s12036-023-09976-2} {\bibfield  {journal} {\bibinfo
  {journal} {J. Astrophys. Astron.}\ }\textbf {\bibinfo {volume} {44}},\
  \bibinfo {pages} {85} (\bibinfo {year} {2023})},\ \Eprint
  {http://arxiv.org/abs/2208.11560} {arXiv:2208.11560 [astro-ph.CO]}
  \BibitemShut {NoStop}%
\bibitem [{\citenamefont {Bassi}\ \emph {et~al.}(2023)\citenamefont {Bassi},
  \citenamefont {Dinda},\ and\ \citenamefont {Sen}}]{Bassi:2023vaq}%
  \BibitemOpen
  \bibfield  {author} {\bibinfo {author} {\bibfnamefont {A.}~\bibnamefont
  {Bassi}}, \bibinfo {author} {\bibfnamefont {B.~R.}\ \bibnamefont {Dinda}}, \
  and\ \bibinfo {author} {\bibfnamefont {A.~A.}\ \bibnamefont {Sen}},\ }\href
  {\doibase 10.1007/s12036-023-09980-6} {\bibfield  {journal} {\bibinfo
  {journal} {J. Astrophys. Astron.}\ }\textbf {\bibinfo {volume} {44}},\
  \bibinfo {pages} {93} (\bibinfo {year} {2023})},\ \Eprint
  {http://arxiv.org/abs/2306.03875} {arXiv:2306.03875 [astro-ph.CO]}
  \BibitemShut {NoStop}%
\bibitem [{\citenamefont {Silvestri}\ and\ \citenamefont
  {Trodden}(2009)}]{Silvestri:2009hh}%
  \BibitemOpen
  \bibfield  {author} {\bibinfo {author} {\bibfnamefont {A.}~\bibnamefont
  {Silvestri}}\ and\ \bibinfo {author} {\bibfnamefont {M.}~\bibnamefont
  {Trodden}},\ }\href {\doibase 10.1088/0034-4885/72/9/096901} {\bibfield
  {journal} {\bibinfo  {journal} {Rept. Prog. Phys.}\ }\textbf {\bibinfo
  {volume} {72}},\ \bibinfo {pages} {096901} (\bibinfo {year} {2009})},\
  \Eprint {http://arxiv.org/abs/0904.0024} {arXiv:0904.0024 [astro-ph.CO]}
  \BibitemShut {NoStop}%
\bibitem [{\citenamefont {Nojiri}\ and\ \citenamefont
  {Odintsov}(2011)}]{Nojiri:2010wj}%
  \BibitemOpen
  \bibfield  {author} {\bibinfo {author} {\bibfnamefont {S.}~\bibnamefont
  {Nojiri}}\ and\ \bibinfo {author} {\bibfnamefont {S.~D.}\ \bibnamefont
  {Odintsov}},\ }\href {\doibase 10.1016/j.physrep.2011.04.001} {\bibfield
  {journal} {\bibinfo  {journal} {Phys. Rept.}\ }\textbf {\bibinfo {volume}
  {505}},\ \bibinfo {pages} {59} (\bibinfo {year} {2011})},\ \Eprint
  {http://arxiv.org/abs/1011.0544} {arXiv:1011.0544 [gr-qc]} \BibitemShut
  {NoStop}%
\bibitem [{\citenamefont {Nojiri}\ and\ \citenamefont
  {Odintsov}(2006)}]{Nojiri:2006ri}%
  \BibitemOpen
  \bibfield  {author} {\bibinfo {author} {\bibfnamefont {S.}~\bibnamefont
  {Nojiri}}\ and\ \bibinfo {author} {\bibfnamefont {S.~D.}\ \bibnamefont
  {Odintsov}},\ }\href {\doibase 10.1142/S0219887807001928} {\bibfield
  {journal} {\bibinfo  {journal} {eConf}\ }\textbf {\bibinfo {volume}
  {C0602061}},\ \bibinfo {pages} {06} (\bibinfo {year} {2006})},\ \Eprint
  {http://arxiv.org/abs/hep-th/0601213} {arXiv:hep-th/0601213} \BibitemShut
  {NoStop}%
\bibitem [{\citenamefont {Nojiri}\ \emph {et~al.}(2017)\citenamefont {Nojiri},
  \citenamefont {Odintsov},\ and\ \citenamefont {Oikonomou}}]{Nojiri:2017ncd}%
  \BibitemOpen
  \bibfield  {author} {\bibinfo {author} {\bibfnamefont {S.}~\bibnamefont
  {Nojiri}}, \bibinfo {author} {\bibfnamefont {S.~D.}\ \bibnamefont
  {Odintsov}}, \ and\ \bibinfo {author} {\bibfnamefont {V.~K.}\ \bibnamefont
  {Oikonomou}},\ }\href {\doibase 10.1016/j.physrep.2017.06.001} {\bibfield
  {journal} {\bibinfo  {journal} {Phys. Rept.}\ }\textbf {\bibinfo {volume}
  {692}},\ \bibinfo {pages} {1} (\bibinfo {year} {2017})},\ \Eprint
  {http://arxiv.org/abs/1705.11098} {arXiv:1705.11098 [gr-qc]} \BibitemShut
  {NoStop}%
\bibitem [{\citenamefont {de~Haro}\ \emph {et~al.}(2023)\citenamefont
  {de~Haro}, \citenamefont {Nojiri}, \citenamefont {Odintsov}, \citenamefont
  {Oikonomou},\ and\ \citenamefont {Pan}}]{deHaro:2023lbq}%
  \BibitemOpen
  \bibfield  {author} {\bibinfo {author} {\bibfnamefont {J.}~\bibnamefont
  {de~Haro}}, \bibinfo {author} {\bibfnamefont {S.}~\bibnamefont {Nojiri}},
  \bibinfo {author} {\bibfnamefont {S.~D.}\ \bibnamefont {Odintsov}}, \bibinfo
  {author} {\bibfnamefont {V.~K.}\ \bibnamefont {Oikonomou}}, \ and\ \bibinfo
  {author} {\bibfnamefont {S.}~\bibnamefont {Pan}},\ }\href {\doibase
  10.1016/j.physrep.2023.09.003} {\bibfield  {journal} {\bibinfo  {journal}
  {Phys. Rept.}\ }\textbf {\bibinfo {volume} {1034}},\ \bibinfo {pages} {1}
  (\bibinfo {year} {2023})},\ \Eprint {http://arxiv.org/abs/2309.07465}
  {arXiv:2309.07465 [gr-qc]} \BibitemShut {NoStop}%
\bibitem [{\citenamefont {Huterer}\ and\ \citenamefont
  {Shafer}(2018)}]{Huterer:2017buf}%
  \BibitemOpen
  \bibfield  {author} {\bibinfo {author} {\bibfnamefont {D.}~\bibnamefont
  {Huterer}}\ and\ \bibinfo {author} {\bibfnamefont {D.~L.}\ \bibnamefont
  {Shafer}},\ }\href {\doibase 10.1088/1361-6633/aa997e} {\bibfield  {journal}
  {\bibinfo  {journal} {Rept. Prog. Phys.}\ }\textbf {\bibinfo {volume} {81}},\
  \bibinfo {pages} {016901} (\bibinfo {year} {2018})},\ \Eprint
  {http://arxiv.org/abs/1709.01091} {arXiv:1709.01091 [astro-ph.CO]}
  \BibitemShut {NoStop}%
\bibitem [{\citenamefont {Motta}\ \emph {et~al.}(2021)\citenamefont {Motta},
  \citenamefont {Garc\'\i{}a-Aspeitia}, \citenamefont {Hern\'andez-Almada},
  \citenamefont {Maga\~na},\ and\ \citenamefont {Verdugo}}]{Motta:2021hvl}%
  \BibitemOpen
  \bibfield  {author} {\bibinfo {author} {\bibfnamefont {V.}~\bibnamefont
  {Motta}}, \bibinfo {author} {\bibfnamefont {M.~A.}\ \bibnamefont
  {Garc\'\i{}a-Aspeitia}}, \bibinfo {author} {\bibfnamefont {A.}~\bibnamefont
  {Hern\'andez-Almada}}, \bibinfo {author} {\bibfnamefont {J.}~\bibnamefont
  {Maga\~na}}, \ and\ \bibinfo {author} {\bibfnamefont {T.}~\bibnamefont
  {Verdugo}},\ }\href {\doibase 10.3390/universe7060163} {\bibfield  {journal}
  {\bibinfo  {journal} {Universe}\ }\textbf {\bibinfo {volume} {7}},\ \bibinfo
  {pages} {163} (\bibinfo {year} {2021})},\ \Eprint
  {http://arxiv.org/abs/2104.04642} {arXiv:2104.04642 [astro-ph.CO]}
  \BibitemShut {NoStop}%
\bibitem [{\citenamefont {Li}\ \emph {et~al.}(2013)\citenamefont {Li},
  \citenamefont {Li}, \citenamefont {Wang},\ and\ \citenamefont
  {Wang}}]{Li:2012dt}%
  \BibitemOpen
  \bibfield  {author} {\bibinfo {author} {\bibfnamefont {M.}~\bibnamefont
  {Li}}, \bibinfo {author} {\bibfnamefont {X.-D.}\ \bibnamefont {Li}}, \bibinfo
  {author} {\bibfnamefont {S.}~\bibnamefont {Wang}}, \ and\ \bibinfo {author}
  {\bibfnamefont {Y.}~\bibnamefont {Wang}},\ }\href {\doibase
  10.1007/s11467-013-0300-5} {\bibfield  {journal} {\bibinfo  {journal} {Front.
  Phys. (Beijing)}\ }\textbf {\bibinfo {volume} {8}},\ \bibinfo {pages} {828}
  (\bibinfo {year} {2013})},\ \Eprint {http://arxiv.org/abs/1209.0922}
  {arXiv:1209.0922 [astro-ph.CO]} \BibitemShut {NoStop}%
\bibitem [{\citenamefont {Carroll}(2001)}]{Carroll:2000fy}%
  \BibitemOpen
  \bibfield  {author} {\bibinfo {author} {\bibfnamefont {S.~M.}\ \bibnamefont
  {Carroll}},\ }\href {\doibase 10.12942/lrr-2001-1} {\bibfield  {journal}
  {\bibinfo  {journal} {Living Rev. Rel.}\ }\textbf {\bibinfo {volume} {4}},\
  \bibinfo {pages} {1} (\bibinfo {year} {2001})},\ \Eprint
  {http://arxiv.org/abs/astro-ph/0004075} {arXiv:astro-ph/0004075} \BibitemShut
  {NoStop}%
\bibitem [{\citenamefont {Zlatev}\ \emph {et~al.}(1999)\citenamefont {Zlatev},
  \citenamefont {Wang},\ and\ \citenamefont {Steinhardt}}]{Zlatev:1998tr}%
  \BibitemOpen
  \bibfield  {author} {\bibinfo {author} {\bibfnamefont {I.}~\bibnamefont
  {Zlatev}}, \bibinfo {author} {\bibfnamefont {L.-M.}\ \bibnamefont {Wang}}, \
  and\ \bibinfo {author} {\bibfnamefont {P.~J.}\ \bibnamefont {Steinhardt}},\
  }\href {\doibase 10.1103/PhysRevLett.82.896} {\bibfield  {journal} {\bibinfo
  {journal} {Phys. Rev. Lett.}\ }\textbf {\bibinfo {volume} {82}},\ \bibinfo
  {pages} {896} (\bibinfo {year} {1999})},\ \Eprint
  {http://arxiv.org/abs/astro-ph/9807002} {arXiv:astro-ph/9807002} \BibitemShut
  {NoStop}%
\bibitem [{\citenamefont {Sahni}\ and\ \citenamefont
  {Starobinsky}(2000)}]{Sahni:1999gb}%
  \BibitemOpen
  \bibfield  {author} {\bibinfo {author} {\bibfnamefont {V.}~\bibnamefont
  {Sahni}}\ and\ \bibinfo {author} {\bibfnamefont {A.~A.}\ \bibnamefont
  {Starobinsky}},\ }\href {\doibase 10.1142/S0218271800000542} {\bibfield
  {journal} {\bibinfo  {journal} {Int. J. Mod. Phys. D}\ }\textbf {\bibinfo
  {volume} {9}},\ \bibinfo {pages} {373} (\bibinfo {year} {2000})},\ \Eprint
  {http://arxiv.org/abs/astro-ph/9904398} {arXiv:astro-ph/9904398} \BibitemShut
  {NoStop}%
\bibitem [{\citenamefont {Velten}\ \emph {et~al.}(2014)\citenamefont {Velten},
  \citenamefont {vom Marttens},\ and\ \citenamefont
  {Zimdahl}}]{Velten:2014nra}%
  \BibitemOpen
  \bibfield  {author} {\bibinfo {author} {\bibfnamefont {H.}~\bibnamefont
  {Velten}}, \bibinfo {author} {\bibfnamefont {R.}~\bibnamefont {vom
  Marttens}}, \ and\ \bibinfo {author} {\bibfnamefont {W.}~\bibnamefont
  {Zimdahl}},\ }\href {\doibase 10.1140/epjc/s10052-014-3160-4} {\bibfield
  {journal} {\bibinfo  {journal} {Eur. Phys. J. C}\ }\textbf {\bibinfo {volume}
  {74}},\ \bibinfo {pages} {3160} (\bibinfo {year} {2014})},\ \Eprint
  {http://arxiv.org/abs/1410.2509} {arXiv:1410.2509 [astro-ph.CO]} \BibitemShut
  {NoStop}%
\bibitem [{\citenamefont {Malquarti}\ \emph {et~al.}(2003)\citenamefont
  {Malquarti}, \citenamefont {Copeland},\ and\ \citenamefont
  {Liddle}}]{Malquarti:2003hn}%
  \BibitemOpen
  \bibfield  {author} {\bibinfo {author} {\bibfnamefont {M.}~\bibnamefont
  {Malquarti}}, \bibinfo {author} {\bibfnamefont {E.~J.}\ \bibnamefont
  {Copeland}}, \ and\ \bibinfo {author} {\bibfnamefont {A.~R.}\ \bibnamefont
  {Liddle}},\ }\href {\doibase 10.1103/PhysRevD.68.023512} {\bibfield
  {journal} {\bibinfo  {journal} {Phys. Rev. D}\ }\textbf {\bibinfo {volume}
  {68}},\ \bibinfo {pages} {023512} (\bibinfo {year} {2003})},\ \Eprint
  {http://arxiv.org/abs/astro-ph/0304277} {arXiv:astro-ph/0304277} \BibitemShut
  {NoStop}%
\bibitem [{\citenamefont {Di~Valentino}\ \emph
  {et~al.}(2021{\natexlab{a}})\citenamefont {Di~Valentino}, \citenamefont
  {Mena}, \citenamefont {Pan}, \citenamefont {Visinelli}, \citenamefont {Yang},
  \citenamefont {Melchiorri}, \citenamefont {Mota}, \citenamefont {Riess},\
  and\ \citenamefont {Silk}}]{DiValentino:2021izs}%
  \BibitemOpen
  \bibfield  {author} {\bibinfo {author} {\bibfnamefont {E.}~\bibnamefont
  {Di~Valentino}}, \bibinfo {author} {\bibfnamefont {O.}~\bibnamefont {Mena}},
  \bibinfo {author} {\bibfnamefont {S.}~\bibnamefont {Pan}}, \bibinfo {author}
  {\bibfnamefont {L.}~\bibnamefont {Visinelli}}, \bibinfo {author}
  {\bibfnamefont {W.}~\bibnamefont {Yang}}, \bibinfo {author} {\bibfnamefont
  {A.}~\bibnamefont {Melchiorri}}, \bibinfo {author} {\bibfnamefont {D.~F.}\
  \bibnamefont {Mota}}, \bibinfo {author} {\bibfnamefont {A.~G.}\ \bibnamefont
  {Riess}}, \ and\ \bibinfo {author} {\bibfnamefont {J.}~\bibnamefont {Silk}},\
  }\href {\doibase 10.1088/1361-6382/ac086d} {\bibfield  {journal} {\bibinfo
  {journal} {Class. Quant. Grav.}\ }\textbf {\bibinfo {volume} {38}},\ \bibinfo
  {pages} {153001} (\bibinfo {year} {2021}{\natexlab{a}})},\ \Eprint
  {http://arxiv.org/abs/2103.01183} {arXiv:2103.01183 [astro-ph.CO]}
  \BibitemShut {NoStop}%
\bibitem [{\citenamefont {Krishnan}\ \emph {et~al.}(2021)\citenamefont
  {Krishnan}, \citenamefont {Mohayaee}, \citenamefont {Colg\'ain},
  \citenamefont {Sheikh-Jabbari},\ and\ \citenamefont
  {Yin}}]{Krishnan:2021dyb}%
  \BibitemOpen
  \bibfield  {author} {\bibinfo {author} {\bibfnamefont {C.}~\bibnamefont
  {Krishnan}}, \bibinfo {author} {\bibfnamefont {R.}~\bibnamefont {Mohayaee}},
  \bibinfo {author} {\bibfnamefont {E.~O.}\ \bibnamefont {Colg\'ain}}, \bibinfo
  {author} {\bibfnamefont {M.~M.}\ \bibnamefont {Sheikh-Jabbari}}, \ and\
  \bibinfo {author} {\bibfnamefont {L.}~\bibnamefont {Yin}},\ }\href {\doibase
  10.1088/1361-6382/ac1a81} {\bibfield  {journal} {\bibinfo  {journal} {Class.
  Quant. Grav.}\ }\textbf {\bibinfo {volume} {38}},\ \bibinfo {pages} {184001}
  (\bibinfo {year} {2021})},\ \Eprint {http://arxiv.org/abs/2105.09790}
  {arXiv:2105.09790 [astro-ph.CO]} \BibitemShut {NoStop}%
\bibitem [{\citenamefont {Vagnozzi}(2020)}]{Vagnozzi:2019ezj}%
  \BibitemOpen
  \bibfield  {author} {\bibinfo {author} {\bibfnamefont {S.}~\bibnamefont
  {Vagnozzi}},\ }\href {\doibase 10.1103/PhysRevD.102.023518} {\bibfield
  {journal} {\bibinfo  {journal} {Phys. Rev. D}\ }\textbf {\bibinfo {volume}
  {102}},\ \bibinfo {pages} {023518} (\bibinfo {year} {2020})},\ \Eprint
  {http://arxiv.org/abs/1907.07569} {arXiv:1907.07569 [astro-ph.CO]}
  \BibitemShut {NoStop}%
\bibitem [{\citenamefont {Dinda}(2022)}]{Dinda:2021ffa}%
  \BibitemOpen
  \bibfield  {author} {\bibinfo {author} {\bibfnamefont {B.~R.}\ \bibnamefont
  {Dinda}},\ }\href {\doibase 10.1103/PhysRevD.105.063524} {\bibfield
  {journal} {\bibinfo  {journal} {Phys. Rev. D}\ }\textbf {\bibinfo {volume}
  {105}},\ \bibinfo {pages} {063524} (\bibinfo {year} {2022})},\ \Eprint
  {http://arxiv.org/abs/2106.02963} {arXiv:2106.02963 [astro-ph.CO]}
  \BibitemShut {NoStop}%
\bibitem [{\citenamefont {Riess}\ \emph {et~al.}(2016)\citenamefont {Riess}
  \emph {et~al.}}]{Riess:2016jrr}%
  \BibitemOpen
  \bibfield  {author} {\bibinfo {author} {\bibfnamefont {A.~G.}\ \bibnamefont
  {Riess}} \emph {et~al.},\ }\href {\doibase 10.3847/0004-637X/826/1/56}
  {\bibfield  {journal} {\bibinfo  {journal} {Astrophys. J.}\ }\textbf
  {\bibinfo {volume} {826}},\ \bibinfo {pages} {56} (\bibinfo {year} {2016})},\
  \Eprint {http://arxiv.org/abs/1604.01424} {arXiv:1604.01424 [astro-ph.CO]}
  \BibitemShut {NoStop}%
\bibitem [{\citenamefont {Riess}\ \emph {et~al.}(2021)\citenamefont {Riess},
  \citenamefont {Casertano}, \citenamefont {Yuan}, \citenamefont {Bowers},
  \citenamefont {Macri}, \citenamefont {Zinn},\ and\ \citenamefont
  {Scolnic}}]{Riess:2020fzl}%
  \BibitemOpen
  \bibfield  {author} {\bibinfo {author} {\bibfnamefont {A.~G.}\ \bibnamefont
  {Riess}}, \bibinfo {author} {\bibfnamefont {S.}~\bibnamefont {Casertano}},
  \bibinfo {author} {\bibfnamefont {W.}~\bibnamefont {Yuan}}, \bibinfo {author}
  {\bibfnamefont {J.~B.}\ \bibnamefont {Bowers}}, \bibinfo {author}
  {\bibfnamefont {L.}~\bibnamefont {Macri}}, \bibinfo {author} {\bibfnamefont
  {J.~C.}\ \bibnamefont {Zinn}}, \ and\ \bibinfo {author} {\bibfnamefont
  {D.}~\bibnamefont {Scolnic}},\ }\href {\doibase 10.3847/2041-8213/abdbaf}
  {\bibfield  {journal} {\bibinfo  {journal} {Astrophys. J. Lett.}\ }\textbf
  {\bibinfo {volume} {908}},\ \bibinfo {pages} {L6} (\bibinfo {year} {2021})},\
  \Eprint {http://arxiv.org/abs/2012.08534} {arXiv:2012.08534 [astro-ph.CO]}
  \BibitemShut {NoStop}%
\bibitem [{\citenamefont {Di~Valentino}\ \emph
  {et~al.}(2021{\natexlab{b}})\citenamefont {Di~Valentino} \emph
  {et~al.}}]{DiValentino:2020vvd}%
  \BibitemOpen
  \bibfield  {author} {\bibinfo {author} {\bibfnamefont {E.}~\bibnamefont
  {Di~Valentino}} \emph {et~al.},\ }\href {\doibase
  10.1016/j.astropartphys.2021.102604} {\bibfield  {journal} {\bibinfo
  {journal} {Astropart. Phys.}\ }\textbf {\bibinfo {volume} {131}},\ \bibinfo
  {pages} {102604} (\bibinfo {year} {2021}{\natexlab{b}})},\ \Eprint
  {http://arxiv.org/abs/2008.11285} {arXiv:2008.11285 [astro-ph.CO]}
  \BibitemShut {NoStop}%
\bibitem [{\citenamefont {Abdalla}\ \emph {et~al.}(2022)\citenamefont {Abdalla}
  \emph {et~al.}}]{Abdalla:2022yfr}%
  \BibitemOpen
  \bibfield  {author} {\bibinfo {author} {\bibfnamefont {E.}~\bibnamefont
  {Abdalla}} \emph {et~al.},\ }\href {\doibase 10.1016/j.jheap.2022.04.002}
  {\bibfield  {journal} {\bibinfo  {journal} {JHEAp}\ }\textbf {\bibinfo
  {volume} {34}},\ \bibinfo {pages} {49} (\bibinfo {year} {2022})},\ \Eprint
  {http://arxiv.org/abs/2203.06142} {arXiv:2203.06142 [astro-ph.CO]}
  \BibitemShut {NoStop}%
\bibitem [{\citenamefont {Douspis}\ \emph {et~al.}(2018)\citenamefont
  {Douspis}, \citenamefont {Salvati},\ and\ \citenamefont
  {Aghanim}}]{Douspis:2018xlj}%
  \BibitemOpen
  \bibfield  {author} {\bibinfo {author} {\bibfnamefont {M.}~\bibnamefont
  {Douspis}}, \bibinfo {author} {\bibfnamefont {L.}~\bibnamefont {Salvati}}, \
  and\ \bibinfo {author} {\bibfnamefont {N.}~\bibnamefont {Aghanim}},\ }\href
  {\doibase 10.22323/1.335.0037} {\bibfield  {journal} {\bibinfo  {journal}
  {PoS}\ }\textbf {\bibinfo {volume} {EDSU2018}},\ \bibinfo {pages} {037}
  (\bibinfo {year} {2018})},\ \Eprint {http://arxiv.org/abs/1901.05289}
  {arXiv:1901.05289 [astro-ph.CO]} \BibitemShut {NoStop}%
\bibitem [{\citenamefont {Bhattacharyya}\ \emph {et~al.}(2019)\citenamefont
  {Bhattacharyya}, \citenamefont {Alam}, \citenamefont {Pandey}, \citenamefont
  {Das},\ and\ \citenamefont {Pal}}]{Bhattacharyya:2018fwb}%
  \BibitemOpen
  \bibfield  {author} {\bibinfo {author} {\bibfnamefont {A.}~\bibnamefont
  {Bhattacharyya}}, \bibinfo {author} {\bibfnamefont {U.}~\bibnamefont {Alam}},
  \bibinfo {author} {\bibfnamefont {K.~L.}\ \bibnamefont {Pandey}}, \bibinfo
  {author} {\bibfnamefont {S.}~\bibnamefont {Das}}, \ and\ \bibinfo {author}
  {\bibfnamefont {S.}~\bibnamefont {Pal}},\ }\href {\doibase
  10.3847/1538-4357/ab12d6} {\bibfield  {journal} {\bibinfo  {journal}
  {Astrophys. J.}\ }\textbf {\bibinfo {volume} {876}},\ \bibinfo {pages} {143}
  (\bibinfo {year} {2019})},\ \Eprint {http://arxiv.org/abs/1805.04716}
  {arXiv:1805.04716 [astro-ph.CO]} \BibitemShut {NoStop}%
\bibitem [{\citenamefont {Dinda}\ and\ \citenamefont
  {Sen}(2018)}]{Dinda:2016ibo}%
  \BibitemOpen
  \bibfield  {author} {\bibinfo {author} {\bibfnamefont {B.~R.}\ \bibnamefont
  {Dinda}}\ and\ \bibinfo {author} {\bibfnamefont {A.~A.}\ \bibnamefont
  {Sen}},\ }\href {\doibase 10.1103/PhysRevD.97.083506} {\bibfield  {journal}
  {\bibinfo  {journal} {Phys. Rev. D}\ }\textbf {\bibinfo {volume} {97}},\
  \bibinfo {pages} {083506} (\bibinfo {year} {2018})},\ \Eprint
  {http://arxiv.org/abs/1607.05123} {arXiv:1607.05123 [astro-ph.CO]}
  \BibitemShut {NoStop}%
\bibitem [{\citenamefont {Dinda}\ and\ \citenamefont
  {Banerjee}(2024)}]{Dinda:2023mad}%
  \BibitemOpen
  \bibfield  {author} {\bibinfo {author} {\bibfnamefont {B.~R.}\ \bibnamefont
  {Dinda}}\ and\ \bibinfo {author} {\bibfnamefont {N.}~\bibnamefont
  {Banerjee}},\ }\href {\doibase 10.1140/epjc/s10052-024-12547-6} {\bibfield
  {journal} {\bibinfo  {journal} {Eur. Phys. J. C}\ }\textbf {\bibinfo {volume}
  {84}},\ \bibinfo {pages} {177} (\bibinfo {year} {2024})},\ \Eprint
  {http://arxiv.org/abs/2309.10538} {arXiv:2309.10538 [astro-ph.CO]}
  \BibitemShut {NoStop}%
\bibitem [{\citenamefont {Anselmi}\ \emph {et~al.}(2014)\citenamefont
  {Anselmi}, \citenamefont {L\'opez~Nacir},\ and\ \citenamefont
  {Sefusatti}}]{Anselmi:2014nya}%
  \BibitemOpen
  \bibfield  {author} {\bibinfo {author} {\bibfnamefont {S.}~\bibnamefont
  {Anselmi}}, \bibinfo {author} {\bibfnamefont {D.}~\bibnamefont
  {L\'opez~Nacir}}, \ and\ \bibinfo {author} {\bibfnamefont {E.}~\bibnamefont
  {Sefusatti}},\ }\href {\doibase 10.1088/1475-7516/2014/07/013} {\bibfield
  {journal} {\bibinfo  {journal} {JCAP}\ }\textbf {\bibinfo {volume} {07}},\
  \bibinfo {pages} {013} (\bibinfo {year} {2014})},\ \Eprint
  {http://arxiv.org/abs/1402.4269} {arXiv:1402.4269 [astro-ph.CO]} \BibitemShut
  {NoStop}%
\bibitem [{\citenamefont {Chevallier}\ and\ \citenamefont
  {Polarski}(2001)}]{Chevallier:2000qy}%
  \BibitemOpen
  \bibfield  {author} {\bibinfo {author} {\bibfnamefont {M.}~\bibnamefont
  {Chevallier}}\ and\ \bibinfo {author} {\bibfnamefont {D.}~\bibnamefont
  {Polarski}},\ }\href {\doibase 10.1142/S0218271801000822} {\bibfield
  {journal} {\bibinfo  {journal} {Int. J. Mod. Phys. D}\ }\textbf {\bibinfo
  {volume} {10}},\ \bibinfo {pages} {213} (\bibinfo {year} {2001})},\ \Eprint
  {http://arxiv.org/abs/gr-qc/0009008} {arXiv:gr-qc/0009008} \BibitemShut
  {NoStop}%
\bibitem [{\citenamefont {Linder}(2003)}]{Linder:2002et}%
  \BibitemOpen
  \bibfield  {author} {\bibinfo {author} {\bibfnamefont {E.~V.}\ \bibnamefont
  {Linder}},\ }\href {\doibase 10.1103/PhysRevLett.90.091301} {\bibfield
  {journal} {\bibinfo  {journal} {Phys. Rev. Lett.}\ }\textbf {\bibinfo
  {volume} {90}},\ \bibinfo {pages} {091301} (\bibinfo {year} {2003})},\
  \Eprint {http://arxiv.org/abs/astro-ph/0208512} {arXiv:astro-ph/0208512}
  \BibitemShut {NoStop}%
\bibitem [{\citenamefont {Barboza}\ and\ \citenamefont
  {Alcaniz}(2008)}]{Barboza:2008rh}%
  \BibitemOpen
  \bibfield  {author} {\bibinfo {author} {\bibfnamefont {E.~M.}\ \bibnamefont
  {Barboza}, \bibfnamefont {Jr.}}\ and\ \bibinfo {author} {\bibfnamefont
  {J.~S.}\ \bibnamefont {Alcaniz}},\ }\href {\doibase
  10.1016/j.physletb.2008.08.012} {\bibfield  {journal} {\bibinfo  {journal}
  {Phys. Lett. B}\ }\textbf {\bibinfo {volume} {666}},\ \bibinfo {pages} {415}
  (\bibinfo {year} {2008})},\ \Eprint {http://arxiv.org/abs/0805.1713}
  {arXiv:0805.1713 [astro-ph]} \BibitemShut {NoStop}%
\bibitem [{\citenamefont {Banerjee}\ \emph {et~al.}(2021)\citenamefont
  {Banerjee}, \citenamefont {Cai}, \citenamefont {Heisenberg}, \citenamefont
  {Colg\'ain}, \citenamefont {Sheikh-Jabbari},\ and\ \citenamefont
  {Yang}}]{Banerjee:2020xcn}%
  \BibitemOpen
  \bibfield  {author} {\bibinfo {author} {\bibfnamefont {A.}~\bibnamefont
  {Banerjee}}, \bibinfo {author} {\bibfnamefont {H.}~\bibnamefont {Cai}},
  \bibinfo {author} {\bibfnamefont {L.}~\bibnamefont {Heisenberg}}, \bibinfo
  {author} {\bibfnamefont {E.~O.}\ \bibnamefont {Colg\'ain}}, \bibinfo {author}
  {\bibfnamefont {M.~M.}\ \bibnamefont {Sheikh-Jabbari}}, \ and\ \bibinfo
  {author} {\bibfnamefont {T.}~\bibnamefont {Yang}},\ }\href {\doibase
  10.1103/PhysRevD.103.L081305} {\bibfield  {journal} {\bibinfo  {journal}
  {Phys. Rev. D}\ }\textbf {\bibinfo {volume} {103}},\ \bibinfo {pages}
  {L081305} (\bibinfo {year} {2021})},\ \Eprint
  {http://arxiv.org/abs/2006.00244} {arXiv:2006.00244 [astro-ph.CO]}
  \BibitemShut {NoStop}%
\bibitem [{\citenamefont {Mehrabi}\ and\ \citenamefont
  {Vazirnia}(2022)}]{Mehrabi:2022ywh}%
  \BibitemOpen
  \bibfield  {author} {\bibinfo {author} {\bibfnamefont {A.}~\bibnamefont
  {Mehrabi}}\ and\ \bibinfo {author} {\bibfnamefont {M.}~\bibnamefont
  {Vazirnia}},\ }\href {\doibase 10.3847/1538-4357/ac6fda} {\bibfield
  {journal} {\bibinfo  {journal} {Astrophys. J.}\ }\textbf {\bibinfo {volume}
  {932}},\ \bibinfo {pages} {121} (\bibinfo {year} {2022})}\BibitemShut
  {NoStop}%
\bibitem [{\citenamefont {Alberto~Vazquez}\ \emph {et~al.}(2012)\citenamefont
  {Alberto~Vazquez}, \citenamefont {Bridges}, \citenamefont {Hobson},\ and\
  \citenamefont {Lasenby}}]{AlbertoVazquez:2012ofj}%
  \BibitemOpen
  \bibfield  {author} {\bibinfo {author} {\bibfnamefont {J.}~\bibnamefont
  {Alberto~Vazquez}}, \bibinfo {author} {\bibfnamefont {M.}~\bibnamefont
  {Bridges}}, \bibinfo {author} {\bibfnamefont {M.~P.}\ \bibnamefont {Hobson}},
  \ and\ \bibinfo {author} {\bibfnamefont {A.~N.}\ \bibnamefont {Lasenby}},\
  }\href {\doibase 10.1088/1475-7516/2012/09/020} {\bibfield  {journal}
  {\bibinfo  {journal} {JCAP}\ }\textbf {\bibinfo {volume} {09}},\ \bibinfo
  {pages} {020} (\bibinfo {year} {2012})},\ \Eprint
  {http://arxiv.org/abs/1205.0847} {arXiv:1205.0847 [astro-ph.CO]} \BibitemShut
  {NoStop}%
\bibitem [{\citenamefont {Liu}\ \emph {et~al.}(2019)\citenamefont {Liu},
  \citenamefont {Qin}, \citenamefont {Zhang}, \citenamefont {Zhang},\ and\
  \citenamefont {Yu}}]{Liu:2015mkm}%
  \BibitemOpen
  \bibfield  {author} {\bibinfo {author} {\bibfnamefont {Z.-E.}\ \bibnamefont
  {Liu}}, \bibinfo {author} {\bibfnamefont {H.-F.}\ \bibnamefont {Qin}},
  \bibinfo {author} {\bibfnamefont {J.}~\bibnamefont {Zhang}}, \bibinfo
  {author} {\bibfnamefont {T.-J.}\ \bibnamefont {Zhang}}, \ and\ \bibinfo
  {author} {\bibfnamefont {H.-R.}\ \bibnamefont {Yu}},\ }\href {\doibase
  10.1016/j.dark.2019.100379} {\bibfield  {journal} {\bibinfo  {journal} {Phys.
  Dark Univ.}\ }\textbf {\bibinfo {volume} {26}},\ \bibinfo {pages} {100379}
  (\bibinfo {year} {2019})},\ \Eprint {http://arxiv.org/abs/1501.02971}
  {arXiv:1501.02971 [astro-ph.CO]} \BibitemShut {NoStop}%
\bibitem [{\citenamefont {Gerardi}\ \emph {et~al.}(2019)\citenamefont
  {Gerardi}, \citenamefont {Martinelli},\ and\ \citenamefont
  {Silvestri}}]{Gerardi:2019obr}%
  \BibitemOpen
  \bibfield  {author} {\bibinfo {author} {\bibfnamefont {F.}~\bibnamefont
  {Gerardi}}, \bibinfo {author} {\bibfnamefont {M.}~\bibnamefont {Martinelli}},
  \ and\ \bibinfo {author} {\bibfnamefont {A.}~\bibnamefont {Silvestri}},\
  }\href {\doibase 10.1088/1475-7516/2019/07/042} {\bibfield  {journal}
  {\bibinfo  {journal} {JCAP}\ }\textbf {\bibinfo {volume} {07}},\ \bibinfo
  {pages} {042} (\bibinfo {year} {2019})},\ \Eprint
  {http://arxiv.org/abs/1902.09423} {arXiv:1902.09423 [astro-ph.CO]}
  \BibitemShut {NoStop}%
\bibitem [{\citenamefont {Bonilla}\ \emph {et~al.}(2021)\citenamefont
  {Bonilla}, \citenamefont {Kumar},\ and\ \citenamefont
  {Nunes}}]{Bonilla:2020wbn}%
  \BibitemOpen
  \bibfield  {author} {\bibinfo {author} {\bibfnamefont {A.}~\bibnamefont
  {Bonilla}}, \bibinfo {author} {\bibfnamefont {S.}~\bibnamefont {Kumar}}, \
  and\ \bibinfo {author} {\bibfnamefont {R.~C.}\ \bibnamefont {Nunes}},\ }\href
  {\doibase 10.1140/epjc/s10052-021-08925-z} {\bibfield  {journal} {\bibinfo
  {journal} {Eur. Phys. J. C}\ }\textbf {\bibinfo {volume} {81}},\ \bibinfo
  {pages} {127} (\bibinfo {year} {2021})},\ \Eprint
  {http://arxiv.org/abs/2011.07140} {arXiv:2011.07140 [astro-ph.CO]}
  \BibitemShut {NoStop}%
\bibitem [{\citenamefont {Sahni}\ and\ \citenamefont
  {Starobinsky}(2006)}]{Sahni:2006pa}%
  \BibitemOpen
  \bibfield  {author} {\bibinfo {author} {\bibfnamefont {V.}~\bibnamefont
  {Sahni}}\ and\ \bibinfo {author} {\bibfnamefont {A.}~\bibnamefont
  {Starobinsky}},\ }\href {\doibase 10.1142/S0218271806009704} {\bibfield
  {journal} {\bibinfo  {journal} {Int. J. Mod. Phys. D}\ }\textbf {\bibinfo
  {volume} {15}},\ \bibinfo {pages} {2105} (\bibinfo {year} {2006})},\ \Eprint
  {http://arxiv.org/abs/astro-ph/0610026} {arXiv:astro-ph/0610026} \BibitemShut
  {NoStop}%
\bibitem [{\citenamefont {Mukherjee}\ and\ \citenamefont
  {Banerjee}(2021)}]{Mukherjee:2020ytg}%
  \BibitemOpen
  \bibfield  {author} {\bibinfo {author} {\bibfnamefont {P.}~\bibnamefont
  {Mukherjee}}\ and\ \bibinfo {author} {\bibfnamefont {N.}~\bibnamefont
  {Banerjee}},\ }\href {\doibase 10.1140/epjc/s10052-021-08830-5} {\bibfield
  {journal} {\bibinfo  {journal} {Eur. Phys. J. C}\ }\textbf {\bibinfo {volume}
  {81}},\ \bibinfo {pages} {36} (\bibinfo {year} {2021})},\ \Eprint
  {http://arxiv.org/abs/2007.10124} {arXiv:2007.10124 [astro-ph.CO]}
  \BibitemShut {NoStop}%
\bibitem [{\citenamefont {Rezaei}\ \emph {et~al.}(2020)\citenamefont {Rezaei},
  \citenamefont {Pour-Ojaghi},\ and\ \citenamefont {Malekjani}}]{Rezaei_2020}%
  \BibitemOpen
  \bibfield  {author} {\bibinfo {author} {\bibfnamefont {M.}~\bibnamefont
  {Rezaei}}, \bibinfo {author} {\bibfnamefont {S.}~\bibnamefont {Pour-Ojaghi}},
  \ and\ \bibinfo {author} {\bibfnamefont {M.}~\bibnamefont {Malekjani}},\
  }\href {\doibase 10.3847/1538-4357/aba517} {\bibfield  {journal} {\bibinfo
  {journal} {The Astrophysical Journal}\ }\textbf {\bibinfo {volume} {900}},\
  \bibinfo {pages} {70} (\bibinfo {year} {2020})}\BibitemShut {NoStop}%
\bibitem [{\citenamefont {Capozziello}\ and\ \citenamefont
  {Agostino}(2022)}]{Capozziello:2022uak}%
  \BibitemOpen
  \bibfield  {author} {\bibinfo {author} {\bibfnamefont {S.}~\bibnamefont
  {Capozziello}}\ and\ \bibinfo {author} {\bibfnamefont {R.~D.~.}\ \bibnamefont
  {Agostino}},\ }\href@noop {} {\bibfield  {journal} {\bibinfo  {journal}
  {Frascati Phys. Ser.}\ }\textbf {\bibinfo {volume} {74}},\ \bibinfo {pages}
  {193} (\bibinfo {year} {2022})},\ \Eprint {http://arxiv.org/abs/2211.17194}
  {arXiv:2211.17194 [astro-ph.CO]} \BibitemShut {NoStop}%
\bibitem [{\citenamefont {Luongo}\ \emph {et~al.}(2016)\citenamefont {Luongo},
  \citenamefont {Pisani},\ and\ \citenamefont {Troisi}}]{Luongo:2015zgq}%
  \BibitemOpen
  \bibfield  {author} {\bibinfo {author} {\bibfnamefont {O.}~\bibnamefont
  {Luongo}}, \bibinfo {author} {\bibfnamefont {G.~B.}\ \bibnamefont {Pisani}},
  \ and\ \bibinfo {author} {\bibfnamefont {A.}~\bibnamefont {Troisi}},\ }\href
  {\doibase 10.1142/S0218271817500158} {\bibfield  {journal} {\bibinfo
  {journal} {Int. J. Mod. Phys. D}\ }\textbf {\bibinfo {volume} {26}},\
  \bibinfo {pages} {1750015} (\bibinfo {year} {2016})},\ \Eprint
  {http://arxiv.org/abs/1512.07076} {arXiv:1512.07076 [gr-qc]} \BibitemShut
  {NoStop}%
\bibitem [{\citenamefont {Wang}\ \emph {et~al.}(2009)\citenamefont {Wang},
  \citenamefont {Dai},\ and\ \citenamefont {Qi}}]{Wang_2009}%
  \BibitemOpen
  \bibfield  {author} {\bibinfo {author} {\bibfnamefont {F.~Y.}\ \bibnamefont
  {Wang}}, \bibinfo {author} {\bibfnamefont {Z.~G.}\ \bibnamefont {Dai}}, \
  and\ \bibinfo {author} {\bibfnamefont {S.}~\bibnamefont {Qi}},\ }\href
  {\doibase 10.1051/0004-6361/200911998} {\bibfield  {journal} {\bibinfo
  {journal} {Astronomy and Astrophysics}\ }\textbf {\bibinfo {volume} {507}},\
  \bibinfo {pages} {53–59} (\bibinfo {year} {2009})}\BibitemShut {NoStop}%
\bibitem [{\citenamefont {Dinda}(2019{\natexlab{a}})}]{Dinda:2019mev}%
  \BibitemOpen
  \bibfield  {author} {\bibinfo {author} {\bibfnamefont {B.~R.}\ \bibnamefont
  {Dinda}},\ }\href {\doibase 10.1103/PhysRevD.100.043528} {\bibfield
  {journal} {\bibinfo  {journal} {Phys. Rev. D}\ }\textbf {\bibinfo {volume}
  {100}},\ \bibinfo {pages} {043528} (\bibinfo {year} {2019}{\natexlab{a}})},\
  \Eprint {http://arxiv.org/abs/1904.10418} {arXiv:1904.10418 [astro-ph.CO]}
  \BibitemShut {NoStop}%
\bibitem [{\citenamefont {Raveri}\ \emph {et~al.}(2023)\citenamefont {Raveri},
  \citenamefont {Pogosian}, \citenamefont {Martinelli}, \citenamefont {Koyama},
  \citenamefont {Silvestri},\ and\ \citenamefont {Zhao}}]{Raveri:2021dbu}%
  \BibitemOpen
  \bibfield  {author} {\bibinfo {author} {\bibfnamefont {M.}~\bibnamefont
  {Raveri}}, \bibinfo {author} {\bibfnamefont {L.}~\bibnamefont {Pogosian}},
  \bibinfo {author} {\bibfnamefont {M.}~\bibnamefont {Martinelli}}, \bibinfo
  {author} {\bibfnamefont {K.}~\bibnamefont {Koyama}}, \bibinfo {author}
  {\bibfnamefont {A.}~\bibnamefont {Silvestri}}, \ and\ \bibinfo {author}
  {\bibfnamefont {G.-B.}\ \bibnamefont {Zhao}},\ }\href {\doibase
  10.1088/1475-7516/2023/02/061} {\bibfield  {journal} {\bibinfo  {journal}
  {JCAP}\ }\textbf {\bibinfo {volume} {02}},\ \bibinfo {pages} {061} (\bibinfo
  {year} {2023})},\ \Eprint {http://arxiv.org/abs/2107.12990} {arXiv:2107.12990
  [astro-ph.CO]} \BibitemShut {NoStop}%
\bibitem [{\citenamefont {Pogosian}\ \emph {et~al.}(2022)\citenamefont
  {Pogosian}, \citenamefont {Raveri}, \citenamefont {Koyama}, \citenamefont
  {Martinelli}, \citenamefont {Silvestri}, \citenamefont {Zhao}, \citenamefont
  {Li}, \citenamefont {Peirone},\ and\ \citenamefont
  {Zucca}}]{Pogosian:2021mcs}%
  \BibitemOpen
  \bibfield  {author} {\bibinfo {author} {\bibfnamefont {L.}~\bibnamefont
  {Pogosian}}, \bibinfo {author} {\bibfnamefont {M.}~\bibnamefont {Raveri}},
  \bibinfo {author} {\bibfnamefont {K.}~\bibnamefont {Koyama}}, \bibinfo
  {author} {\bibfnamefont {M.}~\bibnamefont {Martinelli}}, \bibinfo {author}
  {\bibfnamefont {A.}~\bibnamefont {Silvestri}}, \bibinfo {author}
  {\bibfnamefont {G.-B.}\ \bibnamefont {Zhao}}, \bibinfo {author}
  {\bibfnamefont {J.}~\bibnamefont {Li}}, \bibinfo {author} {\bibfnamefont
  {S.}~\bibnamefont {Peirone}}, \ and\ \bibinfo {author} {\bibfnamefont
  {A.}~\bibnamefont {Zucca}},\ }\href {\doibase 10.1038/s41550-022-01808-7}
  {\bibfield  {journal} {\bibinfo  {journal} {Nature Astron.}\ }\textbf
  {\bibinfo {volume} {6}},\ \bibinfo {pages} {1484} (\bibinfo {year} {2022})},\
  \Eprint {http://arxiv.org/abs/2107.12992} {arXiv:2107.12992 [astro-ph.CO]}
  \BibitemShut {NoStop}%
\bibitem [{\citenamefont {Mu}\ \emph {et~al.}(2023)\citenamefont {Mu},
  \citenamefont {Li},\ and\ \citenamefont {Xu}}]{Mu:2023zct}%
  \BibitemOpen
  \bibfield  {author} {\bibinfo {author} {\bibfnamefont {Y.}~\bibnamefont
  {Mu}}, \bibinfo {author} {\bibfnamefont {E.-K.}\ \bibnamefont {Li}}, \ and\
  \bibinfo {author} {\bibfnamefont {L.}~\bibnamefont {Xu}},\ }\href {\doibase
  10.1088/1475-7516/2023/06/022} {\bibfield  {journal} {\bibinfo  {journal}
  {JCAP}\ }\textbf {\bibinfo {volume} {06}},\ \bibinfo {pages} {022} (\bibinfo
  {year} {2023})},\ \Eprint {http://arxiv.org/abs/2302.09777} {arXiv:2302.09777
  [astro-ph.CO]} \BibitemShut {NoStop}%
\bibitem [{\citenamefont {Ruiz-Zapatero}\ \emph {et~al.}(2022)\citenamefont
  {Ruiz-Zapatero}, \citenamefont {Garc\'\i{}a-Garc\'\i{}a}, \citenamefont
  {Alonso}, \citenamefont {Ferreira},\ and\ \citenamefont
  {Grumitt}}]{Ruiz-Zapatero:2022zpx}%
  \BibitemOpen
  \bibfield  {author} {\bibinfo {author} {\bibfnamefont {J.}~\bibnamefont
  {Ruiz-Zapatero}}, \bibinfo {author} {\bibfnamefont {C.}~\bibnamefont
  {Garc\'\i{}a-Garc\'\i{}a}}, \bibinfo {author} {\bibfnamefont
  {D.}~\bibnamefont {Alonso}}, \bibinfo {author} {\bibfnamefont {P.~G.}\
  \bibnamefont {Ferreira}}, \ and\ \bibinfo {author} {\bibfnamefont {R.~D.~P.}\
  \bibnamefont {Grumitt}},\ }\href {\doibase 10.1093/mnras/stac431} {\bibfield
  {journal} {\bibinfo  {journal} {Mon. Not. Roy. Astron. Soc.}\ }\textbf
  {\bibinfo {volume} {512}},\ \bibinfo {pages} {1967} (\bibinfo {year}
  {2022})},\ \Eprint {http://arxiv.org/abs/2201.07025} {arXiv:2201.07025
  [astro-ph.CO]} \BibitemShut {NoStop}%
\bibitem [{\citenamefont {Bernardo}\ \emph {et~al.}(2022)\citenamefont
  {Bernardo}, \citenamefont {Grand\'on}, \citenamefont {Said~Levi},\ and\
  \citenamefont {C\'ardenas}}]{Bernardo:2021cxi}%
  \BibitemOpen
  \bibfield  {author} {\bibinfo {author} {\bibfnamefont {R.~C.}\ \bibnamefont
  {Bernardo}}, \bibinfo {author} {\bibfnamefont {D.}~\bibnamefont {Grand\'on}},
  \bibinfo {author} {\bibfnamefont {J.}~\bibnamefont {Said~Levi}}, \ and\
  \bibinfo {author} {\bibfnamefont {V.~H.}\ \bibnamefont {C\'ardenas}},\ }\href
  {\doibase 10.1016/j.dark.2022.101017} {\bibfield  {journal} {\bibinfo
  {journal} {Phys. Dark Univ.}\ }\textbf {\bibinfo {volume} {36}},\ \bibinfo
  {pages} {101017} (\bibinfo {year} {2022})},\ \Eprint
  {http://arxiv.org/abs/2111.08289} {arXiv:2111.08289 [astro-ph.CO]}
  \BibitemShut {NoStop}%
\bibitem [{\citenamefont {Calder\'on}\ \emph {et~al.}(2022)\citenamefont
  {Calder\'on}, \citenamefont {L'Huillier}, \citenamefont {Polarski},
  \citenamefont {Shafieloo},\ and\ \citenamefont
  {Starobinsky}}]{Calderon:2022cfj}%
  \BibitemOpen
  \bibfield  {author} {\bibinfo {author} {\bibfnamefont {R.}~\bibnamefont
  {Calder\'on}}, \bibinfo {author} {\bibfnamefont {B.}~\bibnamefont
  {L'Huillier}}, \bibinfo {author} {\bibfnamefont {D.}~\bibnamefont
  {Polarski}}, \bibinfo {author} {\bibfnamefont {A.}~\bibnamefont {Shafieloo}},
  \ and\ \bibinfo {author} {\bibfnamefont {A.~A.}\ \bibnamefont
  {Starobinsky}},\ }\href {\doibase 10.1103/PhysRevD.106.083513} {\bibfield
  {journal} {\bibinfo  {journal} {Phys. Rev. D}\ }\textbf {\bibinfo {volume}
  {106}},\ \bibinfo {pages} {083513} (\bibinfo {year} {2022})},\ \Eprint
  {http://arxiv.org/abs/2206.13820} {arXiv:2206.13820 [astro-ph.CO]}
  \BibitemShut {NoStop}%
\bibitem [{\citenamefont {Perenon}\ \emph {et~al.}(2022)\citenamefont
  {Perenon}, \citenamefont {Martinelli}, \citenamefont {Maartens},
  \citenamefont {Camera},\ and\ \citenamefont {Clarkson}}]{Perenon:2022fgw}%
  \BibitemOpen
  \bibfield  {author} {\bibinfo {author} {\bibfnamefont {L.}~\bibnamefont
  {Perenon}}, \bibinfo {author} {\bibfnamefont {M.}~\bibnamefont {Martinelli}},
  \bibinfo {author} {\bibfnamefont {R.}~\bibnamefont {Maartens}}, \bibinfo
  {author} {\bibfnamefont {S.}~\bibnamefont {Camera}}, \ and\ \bibinfo {author}
  {\bibfnamefont {C.}~\bibnamefont {Clarkson}},\ }\href {\doibase
  10.1016/j.dark.2022.101119} {\bibfield  {journal} {\bibinfo  {journal} {Phys.
  Dark Univ.}\ }\textbf {\bibinfo {volume} {37}},\ \bibinfo {pages} {101119}
  (\bibinfo {year} {2022})},\ \Eprint {http://arxiv.org/abs/2206.12375}
  {arXiv:2206.12375 [astro-ph.CO]} \BibitemShut {NoStop}%
\bibitem [{\citenamefont {Holsclaw}\ \emph {et~al.}(2011)\citenamefont
  {Holsclaw}, \citenamefont {Alam}, \citenamefont {Sansó}, \citenamefont
  {Lee}, \citenamefont {Heitmann}, \citenamefont {Habib},\ and\ \citenamefont
  {Higdon}}]{Holsclaw_2011}%
  \BibitemOpen
  \bibfield  {author} {\bibinfo {author} {\bibfnamefont {T.}~\bibnamefont
  {Holsclaw}}, \bibinfo {author} {\bibfnamefont {U.}~\bibnamefont {Alam}},
  \bibinfo {author} {\bibfnamefont {B.}~\bibnamefont {Sansó}}, \bibinfo
  {author} {\bibfnamefont {H.}~\bibnamefont {Lee}}, \bibinfo {author}
  {\bibfnamefont {K.}~\bibnamefont {Heitmann}}, \bibinfo {author}
  {\bibfnamefont {S.}~\bibnamefont {Habib}}, \ and\ \bibinfo {author}
  {\bibfnamefont {D.}~\bibnamefont {Higdon}},\ }\href {\doibase
  10.1103/physrevd.84.083501} {\bibfield  {journal} {\bibinfo  {journal}
  {Physical Review D}\ }\textbf {\bibinfo {volume} {84}} (\bibinfo {year}
  {2011}),\ 10.1103/physrevd.84.083501}\BibitemShut {NoStop}%
\bibitem [{\citenamefont {Lazkoz}\ \emph {et~al.}(2012)\citenamefont {Lazkoz},
  \citenamefont {Salzano},\ and\ \citenamefont {Sendra}}]{Lazkoz_2012}%
  \BibitemOpen
  \bibfield  {author} {\bibinfo {author} {\bibfnamefont {R.}~\bibnamefont
  {Lazkoz}}, \bibinfo {author} {\bibfnamefont {V.}~\bibnamefont {Salzano}}, \
  and\ \bibinfo {author} {\bibfnamefont {I.}~\bibnamefont {Sendra}},\ }\href
  {\doibase 10.1140/epjc/s10052-012-2130-y} {\bibfield  {journal} {\bibinfo
  {journal} {The European Physical Journal C}\ }\textbf {\bibinfo {volume}
  {72}} (\bibinfo {year} {2012}),\ 10.1140/epjc/s10052-012-2130-y}\BibitemShut
  {NoStop}%
\bibitem [{\citenamefont {Wang}\ \emph {et~al.}(2018)\citenamefont {Wang},
  \citenamefont {Pogosian}, \citenamefont {Zhao},\ and\ \citenamefont
  {Zucca}}]{Wang:2018fng}%
  \BibitemOpen
  \bibfield  {author} {\bibinfo {author} {\bibfnamefont {Y.}~\bibnamefont
  {Wang}}, \bibinfo {author} {\bibfnamefont {L.}~\bibnamefont {Pogosian}},
  \bibinfo {author} {\bibfnamefont {G.-B.}\ \bibnamefont {Zhao}}, \ and\
  \bibinfo {author} {\bibfnamefont {A.}~\bibnamefont {Zucca}},\ }\href
  {\doibase 10.3847/2041-8213/aaf238} {\bibfield  {journal} {\bibinfo
  {journal} {Astrophys. J. Lett.}\ }\textbf {\bibinfo {volume} {869}},\
  \bibinfo {pages} {L8} (\bibinfo {year} {2018})},\ \Eprint
  {http://arxiv.org/abs/1807.03772} {arXiv:1807.03772 [astro-ph.CO]}
  \BibitemShut {NoStop}%
\bibitem [{\citenamefont {Teng}\ \emph {et~al.}(2021)\citenamefont {Teng},
  \citenamefont {Lee},\ and\ \citenamefont {Ng}}]{Teng:2021cvy}%
  \BibitemOpen
  \bibfield  {author} {\bibinfo {author} {\bibfnamefont {Y.-P.}\ \bibnamefont
  {Teng}}, \bibinfo {author} {\bibfnamefont {W.}~\bibnamefont {Lee}}, \ and\
  \bibinfo {author} {\bibfnamefont {K.-W.}\ \bibnamefont {Ng}},\ }\href
  {\doibase 10.1103/PhysRevD.104.083519} {\bibfield  {journal} {\bibinfo
  {journal} {Phys. Rev. D}\ }\textbf {\bibinfo {volume} {104}},\ \bibinfo
  {pages} {083519} (\bibinfo {year} {2021})},\ \Eprint
  {http://arxiv.org/abs/2105.02667} {arXiv:2105.02667 [astro-ph.CO]}
  \BibitemShut {NoStop}%
\bibitem [{\citenamefont {Zhang}\ and\ \citenamefont
  {Li}(2018)}]{Zhang:2018gjb}%
  \BibitemOpen
  \bibfield  {author} {\bibinfo {author} {\bibfnamefont {M.-J.}\ \bibnamefont
  {Zhang}}\ and\ \bibinfo {author} {\bibfnamefont {H.}~\bibnamefont {Li}},\
  }\href {\doibase 10.1140/epjc/s10052-018-5953-3} {\bibfield  {journal}
  {\bibinfo  {journal} {Eur. Phys. J. C}\ }\textbf {\bibinfo {volume} {78}},\
  \bibinfo {pages} {460} (\bibinfo {year} {2018})},\ \Eprint
  {http://arxiv.org/abs/1806.02981} {arXiv:1806.02981 [astro-ph.CO]}
  \BibitemShut {NoStop}%
\bibitem [{\citenamefont {Wang}\ \emph {et~al.}(2007)\citenamefont {Wang},
  \citenamefont {Hui}, \citenamefont {May},\ and\ \citenamefont
  {Haiman}}]{Wang:2007fsa}%
  \BibitemOpen
  \bibfield  {author} {\bibinfo {author} {\bibfnamefont {S.}~\bibnamefont
  {Wang}}, \bibinfo {author} {\bibfnamefont {L.}~\bibnamefont {Hui}}, \bibinfo
  {author} {\bibfnamefont {M.}~\bibnamefont {May}}, \ and\ \bibinfo {author}
  {\bibfnamefont {Z.}~\bibnamefont {Haiman}},\ }\href {\doibase
  10.1103/PhysRevD.76.063503} {\bibfield  {journal} {\bibinfo  {journal} {Phys.
  Rev. D}\ }\textbf {\bibinfo {volume} {76}},\ \bibinfo {pages} {063503}
  (\bibinfo {year} {2007})},\ \Eprint {http://arxiv.org/abs/0705.0165}
  {arXiv:0705.0165 [astro-ph]} \BibitemShut {NoStop}%
\bibitem [{\citenamefont {Ruiz}\ and\ \citenamefont
  {Huterer}(2015)}]{Ruiz:2014hma}%
  \BibitemOpen
  \bibfield  {author} {\bibinfo {author} {\bibfnamefont {E.~J.}\ \bibnamefont
  {Ruiz}}\ and\ \bibinfo {author} {\bibfnamefont {D.}~\bibnamefont {Huterer}},\
  }\href {\doibase 10.1103/PhysRevD.91.063009} {\bibfield  {journal} {\bibinfo
  {journal} {Phys. Rev. D}\ }\textbf {\bibinfo {volume} {91}},\ \bibinfo
  {pages} {063009} (\bibinfo {year} {2015})},\ \Eprint
  {http://arxiv.org/abs/1410.5832} {arXiv:1410.5832 [astro-ph.CO]} \BibitemShut
  {NoStop}%
\bibitem [{\citenamefont {Bernal}\ \emph {et~al.}(2016)\citenamefont {Bernal},
  \citenamefont {Verde},\ and\ \citenamefont {Cuesta}}]{Bernal:2015zom}%
  \BibitemOpen
  \bibfield  {author} {\bibinfo {author} {\bibfnamefont {J.~L.}\ \bibnamefont
  {Bernal}}, \bibinfo {author} {\bibfnamefont {L.}~\bibnamefont {Verde}}, \
  and\ \bibinfo {author} {\bibfnamefont {A.~J.}\ \bibnamefont {Cuesta}},\
  }\href {\doibase 10.1088/1475-7516/2016/02/059} {\bibfield  {journal}
  {\bibinfo  {journal} {JCAP}\ }\textbf {\bibinfo {volume} {02}},\ \bibinfo
  {pages} {059} (\bibinfo {year} {2016})},\ \Eprint
  {http://arxiv.org/abs/1511.03049} {arXiv:1511.03049 [astro-ph.CO]}
  \BibitemShut {NoStop}%
\bibitem [{\citenamefont {D'Agostino}\ and\ \citenamefont
  {Nunes}(2023)}]{DAgostino:2023cgx}%
  \BibitemOpen
  \bibfield  {author} {\bibinfo {author} {\bibfnamefont {R.}~\bibnamefont
  {D'Agostino}}\ and\ \bibinfo {author} {\bibfnamefont {R.~C.}\ \bibnamefont
  {Nunes}},\ }\href {\doibase 10.1103/PhysRevD.108.023523} {\bibfield
  {journal} {\bibinfo  {journal} {Phys. Rev. D}\ }\textbf {\bibinfo {volume}
  {108}},\ \bibinfo {pages} {023523} (\bibinfo {year} {2023})},\ \Eprint
  {http://arxiv.org/abs/2307.13464} {arXiv:2307.13464 [astro-ph.CO]}
  \BibitemShut {NoStop}%
\bibitem [{\citenamefont {Avila}\ \emph {et~al.}(2022)\citenamefont {Avila},
  \citenamefont {Bernui}, \citenamefont {Bonilla},\ and\ \citenamefont
  {Nunes}}]{Avila:2022xad}%
  \BibitemOpen
  \bibfield  {author} {\bibinfo {author} {\bibfnamefont {F.}~\bibnamefont
  {Avila}}, \bibinfo {author} {\bibfnamefont {A.}~\bibnamefont {Bernui}},
  \bibinfo {author} {\bibfnamefont {A.}~\bibnamefont {Bonilla}}, \ and\
  \bibinfo {author} {\bibfnamefont {R.~C.}\ \bibnamefont {Nunes}},\ }\href
  {\doibase 10.1140/epjc/s10052-022-10561-0} {\bibfield  {journal} {\bibinfo
  {journal} {Eur. Phys. J. C}\ }\textbf {\bibinfo {volume} {82}},\ \bibinfo
  {pages} {594} (\bibinfo {year} {2022})},\ \Eprint
  {http://arxiv.org/abs/2201.07829} {arXiv:2201.07829 [astro-ph.CO]}
  \BibitemShut {NoStop}%
\bibitem [{\citenamefont {Williams}\ and\ \citenamefont
  {Rasmussen}(1995)}]{williams1995gaussian}%
  \BibitemOpen
  \bibfield  {author} {\bibinfo {author} {\bibfnamefont {C.}~\bibnamefont
  {Williams}}\ and\ \bibinfo {author} {\bibfnamefont {C.}~\bibnamefont
  {Rasmussen}},\ }\href@noop {} {\bibfield  {journal} {\bibinfo  {journal}
  {Advances in neural information processing systems}\ }\textbf {\bibinfo
  {volume} {8}} (\bibinfo {year} {1995})}\BibitemShut {NoStop}%
\bibitem [{\citenamefont {Rasmussen}\ and\ \citenamefont
  {Williams}(2006)}]{GpRasWil}%
  \BibitemOpen
  \bibfield  {author} {\bibinfo {author} {\bibfnamefont {C.~E.}\ \bibnamefont
  {Rasmussen}}\ and\ \bibinfo {author} {\bibfnamefont {C.~K.~I.}\ \bibnamefont
  {Williams}},\ }\href@noop {} {\emph {\bibinfo {title} {Gaussian Processes for
  Machine Learning}}},\ \bibinfo {edition} {2nd}\ ed.\ (\bibinfo  {publisher}
  {The MIT Press},\ \bibinfo {year} {2006})\BibitemShut {NoStop}%
\bibitem [{\citenamefont {Seikel}\ \emph {et~al.}(2012)\citenamefont {Seikel},
  \citenamefont {Clarkson},\ and\ \citenamefont {Smith}}]{Seikel_2012}%
  \BibitemOpen
  \bibfield  {author} {\bibinfo {author} {\bibfnamefont {M.}~\bibnamefont
  {Seikel}}, \bibinfo {author} {\bibfnamefont {C.}~\bibnamefont {Clarkson}}, \
  and\ \bibinfo {author} {\bibfnamefont {M.}~\bibnamefont {Smith}},\ }\href
  {\doibase 10.1088/1475-7516/2012/06/036} {\bibfield  {journal} {\bibinfo
  {journal} {Journal of Cosmology and Astroparticle Physics}\ }\textbf
  {\bibinfo {volume} {2012}},\ \bibinfo {pages} {036} (\bibinfo {year}
  {2012})}\BibitemShut {NoStop}%
\bibitem [{\citenamefont {Shafieloo}\ \emph {et~al.}(2012)\citenamefont
  {Shafieloo}, \citenamefont {Kim},\ and\ \citenamefont
  {Linder}}]{Shafieloo_2012}%
  \BibitemOpen
  \bibfield  {author} {\bibinfo {author} {\bibfnamefont {A.}~\bibnamefont
  {Shafieloo}}, \bibinfo {author} {\bibfnamefont {A.~G.}\ \bibnamefont {Kim}},
  \ and\ \bibinfo {author} {\bibfnamefont {E.~V.}\ \bibnamefont {Linder}},\
  }\href {\doibase 10.1103/physrevd.85.123530} {\bibfield  {journal} {\bibinfo
  {journal} {Physical Review D}\ }\textbf {\bibinfo {volume} {85}} (\bibinfo
  {year} {2012}),\ 10.1103/physrevd.85.123530}\BibitemShut {NoStop}%
\bibitem [{\citenamefont {Hwang}\ \emph {et~al.}(2023)\citenamefont {Hwang},
  \citenamefont {L'Huillier}, \citenamefont {Keeley}, \citenamefont {Jee},\
  and\ \citenamefont {Shafieloo}}]{Hwang:2022hla}%
  \BibitemOpen
  \bibfield  {author} {\bibinfo {author} {\bibfnamefont {S.-g.}\ \bibnamefont
  {Hwang}}, \bibinfo {author} {\bibfnamefont {B.}~\bibnamefont {L'Huillier}},
  \bibinfo {author} {\bibfnamefont {R.~E.}\ \bibnamefont {Keeley}}, \bibinfo
  {author} {\bibfnamefont {M.~J.}\ \bibnamefont {Jee}}, \ and\ \bibinfo
  {author} {\bibfnamefont {A.}~\bibnamefont {Shafieloo}},\ }\href {\doibase
  10.1088/1475-7516/2023/02/014} {\bibfield  {journal} {\bibinfo  {journal}
  {JCAP}\ }\textbf {\bibinfo {volume} {02}},\ \bibinfo {pages} {014} (\bibinfo
  {year} {2023})},\ \Eprint {http://arxiv.org/abs/2206.15081} {arXiv:2206.15081
  [astro-ph.CO]} \BibitemShut {NoStop}%
\bibitem [{\citenamefont {Keeley}\ \emph {et~al.}(2021)\citenamefont {Keeley},
  \citenamefont {Shafieloo}, \citenamefont {Zhao}, \citenamefont {Vazquez},\
  and\ \citenamefont {Koo}}]{Keeley:2020aym}%
  \BibitemOpen
  \bibfield  {author} {\bibinfo {author} {\bibfnamefont {R.~E.}\ \bibnamefont
  {Keeley}}, \bibinfo {author} {\bibfnamefont {A.}~\bibnamefont {Shafieloo}},
  \bibinfo {author} {\bibfnamefont {G.-B.}\ \bibnamefont {Zhao}}, \bibinfo
  {author} {\bibfnamefont {J.~A.}\ \bibnamefont {Vazquez}}, \ and\ \bibinfo
  {author} {\bibfnamefont {H.}~\bibnamefont {Koo}},\ }\href {\doibase
  10.3847/1538-3881/abdd2a} {\bibfield  {journal} {\bibinfo  {journal} {Astron.
  J.}\ }\textbf {\bibinfo {volume} {161}},\ \bibinfo {pages} {151} (\bibinfo
  {year} {2021})},\ \Eprint {http://arxiv.org/abs/2010.03234} {arXiv:2010.03234
  [astro-ph.CO]} \BibitemShut {NoStop}%
\bibitem [{\citenamefont {Dinda}(2023{\natexlab{a}})}]{Dinda:2022vmb}%
  \BibitemOpen
  \bibfield  {author} {\bibinfo {author} {\bibfnamefont {B.~R.}\ \bibnamefont
  {Dinda}},\ }\href {\doibase 10.1142/S0218271823500797} {\bibfield  {journal}
  {\bibinfo  {journal} {Int. J. Mod. Phys. D}\ }\textbf {\bibinfo {volume}
  {32}},\ \bibinfo {pages} {2350079} (\bibinfo {year} {2023}{\natexlab{a}})},\
  \Eprint {http://arxiv.org/abs/2209.14639} {arXiv:2209.14639 [astro-ph.CO]}
  \BibitemShut {NoStop}%
\bibitem [{\citenamefont {Dinda}\ and\ \citenamefont
  {Banerjee}(2023)}]{Dinda:2022jih}%
  \BibitemOpen
  \bibfield  {author} {\bibinfo {author} {\bibfnamefont {B.~R.}\ \bibnamefont
  {Dinda}}\ and\ \bibinfo {author} {\bibfnamefont {N.}~\bibnamefont
  {Banerjee}},\ }\href {\doibase 10.1103/PhysRevD.107.063513} {\bibfield
  {journal} {\bibinfo  {journal} {Phys. Rev. D}\ }\textbf {\bibinfo {volume}
  {107}},\ \bibinfo {pages} {063513} (\bibinfo {year} {2023})},\ \Eprint
  {http://arxiv.org/abs/2208.14740} {arXiv:2208.14740 [astro-ph.CO]}
  \BibitemShut {NoStop}%
\bibitem [{\citenamefont {Perenon}\ \emph {et~al.}(2021)\citenamefont
  {Perenon}, \citenamefont {Martinelli}, \citenamefont {Ili\'c}, \citenamefont
  {Maartens}, \citenamefont {Lochner},\ and\ \citenamefont
  {Clarkson}}]{Perenon:2021uom}%
  \BibitemOpen
  \bibfield  {author} {\bibinfo {author} {\bibfnamefont {L.}~\bibnamefont
  {Perenon}}, \bibinfo {author} {\bibfnamefont {M.}~\bibnamefont {Martinelli}},
  \bibinfo {author} {\bibfnamefont {S.}~\bibnamefont {Ili\'c}}, \bibinfo
  {author} {\bibfnamefont {R.}~\bibnamefont {Maartens}}, \bibinfo {author}
  {\bibfnamefont {M.}~\bibnamefont {Lochner}}, \ and\ \bibinfo {author}
  {\bibfnamefont {C.}~\bibnamefont {Clarkson}},\ }\href {\doibase
  10.1016/j.dark.2021.100898} {\bibfield  {journal} {\bibinfo  {journal} {Phys.
  Dark Univ.}\ }\textbf {\bibinfo {volume} {34}},\ \bibinfo {pages} {100898}
  (\bibinfo {year} {2021})},\ \Eprint {http://arxiv.org/abs/2105.01613}
  {arXiv:2105.01613 [astro-ph.CO]} \BibitemShut {NoStop}%
\bibitem [{\citenamefont {\'O~Colg\'ain}\ and\ \citenamefont
  {Sheikh-Jabbari}(2021)}]{OColgain:2021pyh}%
  \BibitemOpen
  \bibfield  {author} {\bibinfo {author} {\bibfnamefont {E.}~\bibnamefont
  {\'O~Colg\'ain}}\ and\ \bibinfo {author} {\bibfnamefont {M.~M.}\ \bibnamefont
  {Sheikh-Jabbari}},\ }\href {\doibase 10.1140/epjc/s10052-021-09708-2}
  {\bibfield  {journal} {\bibinfo  {journal} {Eur. Phys. J. C}\ }\textbf
  {\bibinfo {volume} {81}},\ \bibinfo {pages} {892} (\bibinfo {year} {2021})},\
  \Eprint {http://arxiv.org/abs/2101.08565} {arXiv:2101.08565 [astro-ph.CO]}
  \BibitemShut {NoStop}%
\bibitem [{\citenamefont {Banerjee}\ \emph
  {et~al.}(2023{\natexlab{a}})\citenamefont {Banerjee}, \citenamefont
  {Mukherjee},\ and\ \citenamefont {Pav\'on}}]{Banerjee:2023evd}%
  \BibitemOpen
  \bibfield  {author} {\bibinfo {author} {\bibfnamefont {N.}~\bibnamefont
  {Banerjee}}, \bibinfo {author} {\bibfnamefont {P.}~\bibnamefont {Mukherjee}},
  \ and\ \bibinfo {author} {\bibfnamefont {D.}~\bibnamefont {Pav\'on}},\ }\href
  {\doibase 10.1088/1475-7516/2023/11/092} {\bibfield  {journal} {\bibinfo
  {journal} {JCAP}\ }\textbf {\bibinfo {volume} {11}},\ \bibinfo {pages} {092}
  (\bibinfo {year} {2023}{\natexlab{a}})},\ \Eprint
  {http://arxiv.org/abs/2309.12298} {arXiv:2309.12298 [astro-ph.CO]}
  \BibitemShut {NoStop}%
\bibitem [{\citenamefont {Mukherjee}\ \emph {et~al.}(2024)\citenamefont
  {Mukherjee}, \citenamefont {Shah}, \citenamefont {Bhaumik},\ and\
  \citenamefont {Pal}}]{Mukherjee:2023lqr}%
  \BibitemOpen
  \bibfield  {author} {\bibinfo {author} {\bibfnamefont {P.}~\bibnamefont
  {Mukherjee}}, \bibinfo {author} {\bibfnamefont {R.}~\bibnamefont {Shah}},
  \bibinfo {author} {\bibfnamefont {A.}~\bibnamefont {Bhaumik}}, \ and\
  \bibinfo {author} {\bibfnamefont {S.}~\bibnamefont {Pal}},\ }\href {\doibase
  10.3847/1538-4357/ad055f} {\bibfield  {journal} {\bibinfo  {journal}
  {Astrophys. J.}\ }\textbf {\bibinfo {volume} {960}},\ \bibinfo {pages} {61}
  (\bibinfo {year} {2024})},\ \Eprint {http://arxiv.org/abs/2303.05169}
  {arXiv:2303.05169 [astro-ph.CO]} \BibitemShut {NoStop}%
\bibitem [{\citenamefont {Banerjee}\ \emph
  {et~al.}(2023{\natexlab{b}})\citenamefont {Banerjee}, \citenamefont
  {Mukherjee},\ and\ \citenamefont {Pav\'on}}]{Banerjee:2023rvg}%
  \BibitemOpen
  \bibfield  {author} {\bibinfo {author} {\bibfnamefont {N.}~\bibnamefont
  {Banerjee}}, \bibinfo {author} {\bibfnamefont {P.}~\bibnamefont {Mukherjee}},
  \ and\ \bibinfo {author} {\bibfnamefont {D.}~\bibnamefont {Pav\'on}},\ }\href
  {\doibase 10.1093/mnras/stad921} {\bibfield  {journal} {\bibinfo  {journal}
  {Mon. Not. Roy. Astron. Soc.}\ }\textbf {\bibinfo {volume} {521}},\ \bibinfo
  {pages} {5473} (\bibinfo {year} {2023}{\natexlab{b}})},\ \Eprint
  {http://arxiv.org/abs/2301.09823} {arXiv:2301.09823 [astro-ph.CO]}
  \BibitemShut {NoStop}%
\bibitem [{\citenamefont {Mukherjee}\ and\ \citenamefont
  {Banerjee}(2022{\natexlab{a}})}]{Mukherjee:2022ujw}%
  \BibitemOpen
  \bibfield  {author} {\bibinfo {author} {\bibfnamefont {P.}~\bibnamefont
  {Mukherjee}}\ and\ \bibinfo {author} {\bibfnamefont {N.}~\bibnamefont
  {Banerjee}},\ }\href {\doibase 10.1103/PhysRevD.105.063516} {\bibfield
  {journal} {\bibinfo  {journal} {Phys. Rev. D}\ }\textbf {\bibinfo {volume}
  {105}},\ \bibinfo {pages} {063516} (\bibinfo {year} {2022}{\natexlab{a}})},\
  \Eprint {http://arxiv.org/abs/2202.07886} {arXiv:2202.07886 [astro-ph.CO]}
  \BibitemShut {NoStop}%
\bibitem [{\citenamefont {Mukherjee}\ and\ \citenamefont
  {Banerjee}(2022{\natexlab{b}})}]{Mukherjee:2020vkx}%
  \BibitemOpen
  \bibfield  {author} {\bibinfo {author} {\bibfnamefont {P.}~\bibnamefont
  {Mukherjee}}\ and\ \bibinfo {author} {\bibfnamefont {N.}~\bibnamefont
  {Banerjee}},\ }\href {\doibase 10.1016/j.dark.2022.100998} {\bibfield
  {journal} {\bibinfo  {journal} {Phys. Dark Univ.}\ }\textbf {\bibinfo
  {volume} {36}},\ \bibinfo {pages} {100998} (\bibinfo {year}
  {2022}{\natexlab{b}})},\ \Eprint {http://arxiv.org/abs/2007.15941}
  {arXiv:2007.15941 [astro-ph.CO]} \BibitemShut {NoStop}%
\bibitem [{\citenamefont {Zheng}\ \emph {et~al.}(2024)\citenamefont {Zheng},
  \citenamefont {Sakr},\ and\ \citenamefont {Amendola}}]{Zheng:2023yco}%
  \BibitemOpen
  \bibfield  {author} {\bibinfo {author} {\bibfnamefont {Z.}~\bibnamefont
  {Zheng}}, \bibinfo {author} {\bibfnamefont {Z.}~\bibnamefont {Sakr}}, \ and\
  \bibinfo {author} {\bibfnamefont {L.}~\bibnamefont {Amendola}},\ }\href
  {\doibase 10.1016/j.physletb.2024.138647} {\bibfield  {journal} {\bibinfo
  {journal} {Phys. Lett. B}\ }\textbf {\bibinfo {volume} {853}},\ \bibinfo
  {pages} {138647} (\bibinfo {year} {2024})},\ \Eprint
  {http://arxiv.org/abs/2312.07436} {arXiv:2312.07436 [astro-ph.CO]}
  \BibitemShut {NoStop}%
\bibitem [{\citenamefont {Oliveira}\ \emph {et~al.}(2024)\citenamefont
  {Oliveira}, \citenamefont {Avila}, \citenamefont {Bernui}, \citenamefont
  {Bonilla},\ and\ \citenamefont {Nunes}}]{Oliveira:2023uid}%
  \BibitemOpen
  \bibfield  {author} {\bibinfo {author} {\bibfnamefont {F.}~\bibnamefont
  {Oliveira}}, \bibinfo {author} {\bibfnamefont {F.}~\bibnamefont {Avila}},
  \bibinfo {author} {\bibfnamefont {A.}~\bibnamefont {Bernui}}, \bibinfo
  {author} {\bibfnamefont {A.}~\bibnamefont {Bonilla}}, \ and\ \bibinfo
  {author} {\bibfnamefont {R.~C.}\ \bibnamefont {Nunes}},\ }\href {\doibase
  10.1140/epjc/s10052-024-12953-w} {\bibfield  {journal} {\bibinfo  {journal}
  {Eur. Phys. J. C}\ }\textbf {\bibinfo {volume} {84}},\ \bibinfo {pages} {636}
  (\bibinfo {year} {2024})},\ \Eprint {http://arxiv.org/abs/2311.14216}
  {arXiv:2311.14216 [astro-ph.CO]} \BibitemShut {NoStop}%
\bibitem [{\citenamefont {Dinda}(2019{\natexlab{b}})}]{Dinda:2018ojk}%
  \BibitemOpen
  \bibfield  {author} {\bibinfo {author} {\bibfnamefont {B.~R.}\ \bibnamefont
  {Dinda}},\ }\href {\doibase 10.1007/s12036-019-9584-3} {\bibfield  {journal}
  {\bibinfo  {journal} {J. Astrophys. Astron.}\ }\textbf {\bibinfo {volume}
  {40}},\ \bibinfo {pages} {12} (\bibinfo {year} {2019}{\natexlab{b}})},\
  \Eprint {http://arxiv.org/abs/1804.07953} {arXiv:1804.07953 [astro-ph.CO]}
  \BibitemShut {NoStop}%
\bibitem [{\citenamefont {Dinda}(2023{\natexlab{b}})}]{Dinda:2023kvg}%
  \BibitemOpen
  \bibfield  {author} {\bibinfo {author} {\bibfnamefont {B.~R.}\ \bibnamefont
  {Dinda}},\ }\href@noop {} {\  (\bibinfo {year} {2023}{\natexlab{b}})},\
  \Eprint {http://arxiv.org/abs/2312.01393} {arXiv:2312.01393 [astro-ph.CO]}
  \BibitemShut {NoStop}%
\bibitem [{\citenamefont {Dinda}(2024)}]{Dinda:2023xqx}%
  \BibitemOpen
  \bibfield  {author} {\bibinfo {author} {\bibfnamefont {B.~R.}\ \bibnamefont
  {Dinda}},\ }\href {\doibase 10.1140/epjc/s10052-024-12774-x} {\bibfield
  {journal} {\bibinfo  {journal} {Eur. Phys. J. C}\ }\textbf {\bibinfo {volume}
  {84}},\ \bibinfo {pages} {402} (\bibinfo {year} {2024})},\ \Eprint
  {http://arxiv.org/abs/2311.13498} {arXiv:2311.13498 [astro-ph.CO]}
  \BibitemShut {NoStop}%
\bibitem [{\citenamefont {Moresco}\ \emph {et~al.}(2020)\citenamefont
  {Moresco}, \citenamefont {Jimenez}, \citenamefont {Verde}, \citenamefont
  {Cimatti},\ and\ \citenamefont {Pozzetti}}]{Moresco:2020fbm}%
  \BibitemOpen
  \bibfield  {author} {\bibinfo {author} {\bibfnamefont {M.}~\bibnamefont
  {Moresco}}, \bibinfo {author} {\bibfnamefont {R.}~\bibnamefont {Jimenez}},
  \bibinfo {author} {\bibfnamefont {L.}~\bibnamefont {Verde}}, \bibinfo
  {author} {\bibfnamefont {A.}~\bibnamefont {Cimatti}}, \ and\ \bibinfo
  {author} {\bibfnamefont {L.}~\bibnamefont {Pozzetti}},\ }\href {\doibase
  10.3847/1538-4357/ab9eb0} {\bibfield  {journal} {\bibinfo  {journal}
  {Astrophys. J.}\ }\textbf {\bibinfo {volume} {898}},\ \bibinfo {pages} {82}
  (\bibinfo {year} {2020})},\ \Eprint {http://arxiv.org/abs/2003.07362}
  {arXiv:2003.07362 [astro-ph.GA]} \BibitemShut {NoStop}%
\bibitem [{\citenamefont {Moresco}\ \emph {et~al.}(2012)\citenamefont
  {Moresco}, \citenamefont {Cimatti}, \citenamefont {Jimenez}, \citenamefont
  {Pozzetti}, \citenamefont {Zamorani}, \citenamefont {Bolzonella},
  \citenamefont {Dunlop}, \citenamefont {Lamareille}, \citenamefont {Mignoli},
  \citenamefont {Pearce} \emph {et~al.}}]{moresco2012improved}%
  \BibitemOpen
  \bibfield  {author} {\bibinfo {author} {\bibfnamefont {M.}~\bibnamefont
  {Moresco}}, \bibinfo {author} {\bibfnamefont {A.}~\bibnamefont {Cimatti}},
  \bibinfo {author} {\bibfnamefont {R.}~\bibnamefont {Jimenez}}, \bibinfo
  {author} {\bibfnamefont {L.}~\bibnamefont {Pozzetti}}, \bibinfo {author}
  {\bibfnamefont {G.}~\bibnamefont {Zamorani}}, \bibinfo {author}
  {\bibfnamefont {M.}~\bibnamefont {Bolzonella}}, \bibinfo {author}
  {\bibfnamefont {J.}~\bibnamefont {Dunlop}}, \bibinfo {author} {\bibfnamefont
  {F.}~\bibnamefont {Lamareille}}, \bibinfo {author} {\bibfnamefont
  {M.}~\bibnamefont {Mignoli}}, \bibinfo {author} {\bibfnamefont
  {H.}~\bibnamefont {Pearce}},  \emph {et~al.},\ }\href@noop {} {\bibfield
  {journal} {\bibinfo  {journal} {Journal of Cosmology and Astroparticle
  Physics}\ }\textbf {\bibinfo {volume} {2012}},\ \bibinfo {pages} {006}
  (\bibinfo {year} {2012})}\BibitemShut {NoStop}%
\bibitem [{\citenamefont {Moresco}(2015)}]{Moresco:2015cya}%
  \BibitemOpen
  \bibfield  {author} {\bibinfo {author} {\bibfnamefont {M.}~\bibnamefont
  {Moresco}},\ }\href {\doibase 10.1093/mnrasl/slv037} {\bibfield  {journal}
  {\bibinfo  {journal} {Mon. Not. Roy. Astron. Soc.}\ }\textbf {\bibinfo
  {volume} {450}},\ \bibinfo {pages} {L16} (\bibinfo {year} {2015})},\ \Eprint
  {http://arxiv.org/abs/1503.01116} {arXiv:1503.01116 [astro-ph.CO]}
  \BibitemShut {NoStop}%
\bibitem [{\citenamefont {Moresco}\ \emph {et~al.}(2016)\citenamefont
  {Moresco}, \citenamefont {Pozzetti}, \citenamefont {Cimatti}, \citenamefont
  {Jimenez}, \citenamefont {Maraston}, \citenamefont {Verde}, \citenamefont
  {Thomas}, \citenamefont {Citro}, \citenamefont {Tojeiro},\ and\ \citenamefont
  {Wilkinson}}]{Moresco:2016mzx}%
  \BibitemOpen
  \bibfield  {author} {\bibinfo {author} {\bibfnamefont {M.}~\bibnamefont
  {Moresco}}, \bibinfo {author} {\bibfnamefont {L.}~\bibnamefont {Pozzetti}},
  \bibinfo {author} {\bibfnamefont {A.}~\bibnamefont {Cimatti}}, \bibinfo
  {author} {\bibfnamefont {R.}~\bibnamefont {Jimenez}}, \bibinfo {author}
  {\bibfnamefont {C.}~\bibnamefont {Maraston}}, \bibinfo {author}
  {\bibfnamefont {L.}~\bibnamefont {Verde}}, \bibinfo {author} {\bibfnamefont
  {D.}~\bibnamefont {Thomas}}, \bibinfo {author} {\bibfnamefont
  {A.}~\bibnamefont {Citro}}, \bibinfo {author} {\bibfnamefont
  {R.}~\bibnamefont {Tojeiro}}, \ and\ \bibinfo {author} {\bibfnamefont
  {D.}~\bibnamefont {Wilkinson}},\ }\href {\doibase
  10.1088/1475-7516/2016/05/014} {\bibfield  {journal} {\bibinfo  {journal}
  {JCAP}\ }\textbf {\bibinfo {volume} {05}},\ \bibinfo {pages} {014} (\bibinfo
  {year} {2016})},\ \Eprint {http://arxiv.org/abs/1601.01701} {arXiv:1601.01701
  [astro-ph.CO]} \BibitemShut {NoStop}%
\bibitem [{\citenamefont {Zhang}\ \emph {et~al.}(2014)\citenamefont {Zhang},
  \citenamefont {Zhang}, \citenamefont {Yuan}, \citenamefont {Liu},
  \citenamefont {Zhang},\ and\ \citenamefont {Sun}}]{zhang2014four}%
  \BibitemOpen
  \bibfield  {author} {\bibinfo {author} {\bibfnamefont {C.}~\bibnamefont
  {Zhang}}, \bibinfo {author} {\bibfnamefont {H.}~\bibnamefont {Zhang}},
  \bibinfo {author} {\bibfnamefont {S.}~\bibnamefont {Yuan}}, \bibinfo {author}
  {\bibfnamefont {S.}~\bibnamefont {Liu}}, \bibinfo {author} {\bibfnamefont
  {T.-J.}\ \bibnamefont {Zhang}}, \ and\ \bibinfo {author} {\bibfnamefont
  {Y.-C.}\ \bibnamefont {Sun}},\ }\href@noop {} {\bibfield  {journal} {\bibinfo
   {journal} {Research in Astronomy and Astrophysics}\ }\textbf {\bibinfo
  {volume} {14}},\ \bibinfo {pages} {1221} (\bibinfo {year}
  {2014})}\BibitemShut {NoStop}%
\bibitem [{\citenamefont {Simon}\ \emph {et~al.}(2005)\citenamefont {Simon},
  \citenamefont {Verde},\ and\ \citenamefont {Jimenez}}]{Simon:2004tf}%
  \BibitemOpen
  \bibfield  {author} {\bibinfo {author} {\bibfnamefont {J.}~\bibnamefont
  {Simon}}, \bibinfo {author} {\bibfnamefont {L.}~\bibnamefont {Verde}}, \ and\
  \bibinfo {author} {\bibfnamefont {R.}~\bibnamefont {Jimenez}},\ }\href
  {\doibase 10.1103/PhysRevD.71.123001} {\bibfield  {journal} {\bibinfo
  {journal} {Phys. Rev. D}\ }\textbf {\bibinfo {volume} {71}},\ \bibinfo
  {pages} {123001} (\bibinfo {year} {2005})},\ \Eprint
  {http://arxiv.org/abs/astro-ph/0412269} {arXiv:astro-ph/0412269} \BibitemShut
  {NoStop}%
\bibitem [{\citenamefont {Ratsimbazafy}\ \emph {et~al.}(2017)\citenamefont
  {Ratsimbazafy}, \citenamefont {Loubser}, \citenamefont {Crawford},
  \citenamefont {Cress}, \citenamefont {Bassett}, \citenamefont {Nichol},\ and\
  \citenamefont {V\"ais\"anen}}]{Ratsimbazafy:2017vga}%
  \BibitemOpen
  \bibfield  {author} {\bibinfo {author} {\bibfnamefont {A.~L.}\ \bibnamefont
  {Ratsimbazafy}}, \bibinfo {author} {\bibfnamefont {S.~I.}\ \bibnamefont
  {Loubser}}, \bibinfo {author} {\bibfnamefont {S.~M.}\ \bibnamefont
  {Crawford}}, \bibinfo {author} {\bibfnamefont {C.~M.}\ \bibnamefont {Cress}},
  \bibinfo {author} {\bibfnamefont {B.~A.}\ \bibnamefont {Bassett}}, \bibinfo
  {author} {\bibfnamefont {R.~C.}\ \bibnamefont {Nichol}}, \ and\ \bibinfo
  {author} {\bibfnamefont {P.}~\bibnamefont {V\"ais\"anen}},\ }\href {\doibase
  10.1093/mnras/stx301} {\bibfield  {journal} {\bibinfo  {journal} {Mon. Not.
  Roy. Astron. Soc.}\ }\textbf {\bibinfo {volume} {467}},\ \bibinfo {pages}
  {3239} (\bibinfo {year} {2017})},\ \Eprint {http://arxiv.org/abs/1702.00418}
  {arXiv:1702.00418 [astro-ph.CO]} \BibitemShut {NoStop}%
\bibitem [{\citenamefont {Stern}\ \emph {et~al.}(2010)\citenamefont {Stern},
  \citenamefont {Jimenez}, \citenamefont {Verde}, \citenamefont
  {Kamionkowski},\ and\ \citenamefont {Stanford}}]{stern2010cosmic}%
  \BibitemOpen
  \bibfield  {author} {\bibinfo {author} {\bibfnamefont {D.}~\bibnamefont
  {Stern}}, \bibinfo {author} {\bibfnamefont {R.}~\bibnamefont {Jimenez}},
  \bibinfo {author} {\bibfnamefont {L.}~\bibnamefont {Verde}}, \bibinfo
  {author} {\bibfnamefont {M.}~\bibnamefont {Kamionkowski}}, \ and\ \bibinfo
  {author} {\bibfnamefont {S.~A.}\ \bibnamefont {Stanford}},\ }\href@noop {}
  {\bibfield  {journal} {\bibinfo  {journal} {Journal of Cosmology and
  Astroparticle Physics}\ }\textbf {\bibinfo {volume} {2010}},\ \bibinfo
  {pages} {008} (\bibinfo {year} {2010})}\BibitemShut {NoStop}%
\bibitem [{\citenamefont {Borghi}\ \emph {et~al.}(2022)\citenamefont {Borghi},
  \citenamefont {Moresco},\ and\ \citenamefont {Cimatti}}]{Borghi:2021rft}%
  \BibitemOpen
  \bibfield  {author} {\bibinfo {author} {\bibfnamefont {N.}~\bibnamefont
  {Borghi}}, \bibinfo {author} {\bibfnamefont {M.}~\bibnamefont {Moresco}}, \
  and\ \bibinfo {author} {\bibfnamefont {A.}~\bibnamefont {Cimatti}},\ }\href
  {\doibase 10.3847/2041-8213/ac3fb2} {\bibfield  {journal} {\bibinfo
  {journal} {Astrophys. J. Lett.}\ }\textbf {\bibinfo {volume} {928}},\
  \bibinfo {pages} {L4} (\bibinfo {year} {2022})},\ \Eprint
  {http://arxiv.org/abs/2110.04304} {arXiv:2110.04304 [astro-ph.CO]}
  \BibitemShut {NoStop}%
\bibitem [{\citenamefont {Avila}\ \emph {et~al.}(2021)\citenamefont {Avila},
  \citenamefont {Bernui}, \citenamefont {de~Carvalho},\ and\ \citenamefont
  {Novaes}}]{Avila:2021dqv}%
  \BibitemOpen
  \bibfield  {author} {\bibinfo {author} {\bibfnamefont {F.}~\bibnamefont
  {Avila}}, \bibinfo {author} {\bibfnamefont {A.}~\bibnamefont {Bernui}},
  \bibinfo {author} {\bibfnamefont {E.}~\bibnamefont {de~Carvalho}}, \ and\
  \bibinfo {author} {\bibfnamefont {C.~P.}\ \bibnamefont {Novaes}},\ }\href
  {\doibase 10.1093/mnras/stab1488} {\bibfield  {journal} {\bibinfo  {journal}
  {Mon. Not. Roy. Astron. Soc.}\ }\textbf {\bibinfo {volume} {505}},\ \bibinfo
  {pages} {3404} (\bibinfo {year} {2021})},\ \Eprint
  {http://arxiv.org/abs/2105.10583} {arXiv:2105.10583 [astro-ph.CO]}
  \BibitemShut {NoStop}%
\bibitem [{\citenamefont {Hawkins}\ \emph {et~al.}(2003)\citenamefont {Hawkins}
  \emph {et~al.}}]{Hawkins:2002sg}%
  \BibitemOpen
  \bibfield  {author} {\bibinfo {author} {\bibfnamefont {E.}~\bibnamefont
  {Hawkins}} \emph {et~al.},\ }\href {\doibase
  10.1046/j.1365-2966.2003.07063.x} {\bibfield  {journal} {\bibinfo  {journal}
  {Mon. Not. Roy. Astron. Soc.}\ }\textbf {\bibinfo {volume} {346}},\ \bibinfo
  {pages} {78} (\bibinfo {year} {2003})},\ \Eprint
  {http://arxiv.org/abs/astro-ph/0212375} {arXiv:astro-ph/0212375} \BibitemShut
  {NoStop}%
\bibitem [{\citenamefont {Guzzo}\ \emph {et~al.}(2008)\citenamefont {Guzzo}
  \emph {et~al.}}]{Guzzo:2008ac}%
  \BibitemOpen
  \bibfield  {author} {\bibinfo {author} {\bibfnamefont {L.}~\bibnamefont
  {Guzzo}} \emph {et~al.},\ }\href {\doibase 10.1038/nature06555} {\bibfield
  {journal} {\bibinfo  {journal} {Nature}\ }\textbf {\bibinfo {volume} {451}},\
  \bibinfo {pages} {541} (\bibinfo {year} {2008})},\ \Eprint
  {http://arxiv.org/abs/0802.1944} {arXiv:0802.1944 [astro-ph]} \BibitemShut
  {NoStop}%
\bibitem [{\citenamefont {Blake}\ \emph {et~al.}(2013)\citenamefont {Blake}
  \emph {et~al.}}]{Blake:2013nif}%
  \BibitemOpen
  \bibfield  {author} {\bibinfo {author} {\bibfnamefont {C.}~\bibnamefont
  {Blake}} \emph {et~al.},\ }\href {\doibase 10.1093/mnras/stt1791} {\bibfield
  {journal} {\bibinfo  {journal} {Mon. Not. Roy. Astron. Soc.}\ }\textbf
  {\bibinfo {volume} {436}},\ \bibinfo {pages} {3089} (\bibinfo {year}
  {2013})},\ \Eprint {http://arxiv.org/abs/1309.5556} {arXiv:1309.5556
  [astro-ph.CO]} \BibitemShut {NoStop}%
\bibitem [{\citenamefont {Blake}\ \emph {et~al.}(2011)\citenamefont {Blake},
  \citenamefont {Brough}, \citenamefont {Colless}, \citenamefont {Contreras},
  \citenamefont {Couch}, \citenamefont {Croom}, \citenamefont {Davis},
  \citenamefont {Drinkwater}, \citenamefont {Forster}, \citenamefont {Gilbank},
  \citenamefont {Gladders}, \citenamefont {Glazebrook}, \citenamefont
  {Jelliffe}, \citenamefont {Jurek}, \citenamefont {Li}, \citenamefont
  {Madore}, \citenamefont {Martin}, \citenamefont {Pimbblet}, \citenamefont
  {Poole}, \citenamefont {Pracy}, \citenamefont {Sharp}, \citenamefont
  {Wisnioski}, \citenamefont {Woods}, \citenamefont {Wyder},\ and\
  \citenamefont {Yee}}]{Blake_2011}%
  \BibitemOpen
  \bibfield  {author} {\bibinfo {author} {\bibfnamefont {C.}~\bibnamefont
  {Blake}}, \bibinfo {author} {\bibfnamefont {S.}~\bibnamefont {Brough}},
  \bibinfo {author} {\bibfnamefont {M.}~\bibnamefont {Colless}}, \bibinfo
  {author} {\bibfnamefont {C.}~\bibnamefont {Contreras}}, \bibinfo {author}
  {\bibfnamefont {W.}~\bibnamefont {Couch}}, \bibinfo {author} {\bibfnamefont
  {S.}~\bibnamefont {Croom}}, \bibinfo {author} {\bibfnamefont
  {T.}~\bibnamefont {Davis}}, \bibinfo {author} {\bibfnamefont {M.~J.}\
  \bibnamefont {Drinkwater}}, \bibinfo {author} {\bibfnamefont
  {K.}~\bibnamefont {Forster}}, \bibinfo {author} {\bibfnamefont
  {D.}~\bibnamefont {Gilbank}}, \bibinfo {author} {\bibfnamefont
  {M.}~\bibnamefont {Gladders}}, \bibinfo {author} {\bibfnamefont
  {K.}~\bibnamefont {Glazebrook}}, \bibinfo {author} {\bibfnamefont
  {B.}~\bibnamefont {Jelliffe}}, \bibinfo {author} {\bibfnamefont {R.~J.}\
  \bibnamefont {Jurek}}, \bibinfo {author} {\bibfnamefont {I.-h.}\ \bibnamefont
  {Li}}, \bibinfo {author} {\bibfnamefont {B.}~\bibnamefont {Madore}}, \bibinfo
  {author} {\bibfnamefont {D.~C.}\ \bibnamefont {Martin}}, \bibinfo {author}
  {\bibfnamefont {K.}~\bibnamefont {Pimbblet}}, \bibinfo {author}
  {\bibfnamefont {G.~B.}\ \bibnamefont {Poole}}, \bibinfo {author}
  {\bibfnamefont {M.}~\bibnamefont {Pracy}}, \bibinfo {author} {\bibfnamefont
  {R.}~\bibnamefont {Sharp}}, \bibinfo {author} {\bibfnamefont
  {E.}~\bibnamefont {Wisnioski}}, \bibinfo {author} {\bibfnamefont
  {D.}~\bibnamefont {Woods}}, \bibinfo {author} {\bibfnamefont {T.~K.}\
  \bibnamefont {Wyder}}, \ and\ \bibinfo {author} {\bibfnamefont {H.~K.~C.}\
  \bibnamefont {Yee}},\ }\href {\doibase 10.1111/j.1365-2966.2011.18903.x}
  {\bibfield  {journal} {\bibinfo  {journal} {Monthly Notices of the Royal
  Astronomical Society}\ }\textbf {\bibinfo {volume} {415}},\ \bibinfo {pages}
  {2876–2891} (\bibinfo {year} {2011})}\BibitemShut {NoStop}%
\bibitem [{\citenamefont {Tegmark}\ \emph {et~al.}(2006)\citenamefont {Tegmark}
  \emph {et~al.}}]{SDSS:2006lmn}%
  \BibitemOpen
  \bibfield  {author} {\bibinfo {author} {\bibfnamefont {M.}~\bibnamefont
  {Tegmark}} \emph {et~al.} (\bibinfo {collaboration} {SDSS}),\ }\href
  {\doibase 10.1103/PhysRevD.74.123507} {\bibfield  {journal} {\bibinfo
  {journal} {Phys. Rev. D}\ }\textbf {\bibinfo {volume} {74}},\ \bibinfo
  {pages} {123507} (\bibinfo {year} {2006})},\ \Eprint
  {http://arxiv.org/abs/astro-ph/0608632} {arXiv:astro-ph/0608632} \BibitemShut
  {NoStop}%
\bibitem [{\citenamefont {Ross}\ \emph {et~al.}(2007)\citenamefont {Ross} \emph
  {et~al.}}]{Ross:2006me}%
  \BibitemOpen
  \bibfield  {author} {\bibinfo {author} {\bibfnamefont {N.~P.}\ \bibnamefont
  {Ross}} \emph {et~al.},\ }\href {\doibase 10.1111/j.1365-2966.2007.12289.x}
  {\bibfield  {journal} {\bibinfo  {journal} {Mon. Not. Roy. Astron. Soc.}\
  }\textbf {\bibinfo {volume} {381}},\ \bibinfo {pages} {573} (\bibinfo {year}
  {2007})},\ \Eprint {http://arxiv.org/abs/astro-ph/0612400}
  {arXiv:astro-ph/0612400} \BibitemShut {NoStop}%
\bibitem [{\citenamefont {da~Angela}\ \emph {et~al.}(2008)\citenamefont
  {da~Angela} \emph {et~al.}}]{daAngela:2006mf}%
  \BibitemOpen
  \bibfield  {author} {\bibinfo {author} {\bibfnamefont {J.}~\bibnamefont
  {da~Angela}} \emph {et~al.},\ }\href {\doibase
  10.1111/j.1365-2966.2007.12552.x} {\bibfield  {journal} {\bibinfo  {journal}
  {Mon. Not. Roy. Astron. Soc.}\ }\textbf {\bibinfo {volume} {383}},\ \bibinfo
  {pages} {565} (\bibinfo {year} {2008})},\ \Eprint
  {http://arxiv.org/abs/astro-ph/0612401} {arXiv:astro-ph/0612401} \BibitemShut
  {NoStop}%
\end{thebibliography}%

\end{document}